\documentclass[prd,twocolumn,amssymb,eqsecnum,showpacs]{revtex4}
\usepackage{epsfig}  

\newcommand\egg{\eta^\prime g^* g^*}

\begin{document}

\preprint{DESY 00-093} 
\preprint{YARU-HE-00/05} 

\title{The $\eta^\prime g^* g^*$ Vertex with Arbitrary Gluon Virtualities \\ 
       in the Perturbative QCD Hard Scattering Approach}

\author{Ahmed Ali} 
\affiliation{Deutsches Elektronen Synchrotron DESY, Notkestra\ss e 85,
             22607 Hamburg, Germany}
\email{ali@mail.desy.de}

\author{Alexander Ya. Parkhomenko}
\affiliation{Department of Theoretical Physics, 
             Yaroslavl State University, 
             Sovietskaya 14, 150000 Yaroslavl, Russia} 
\email{parkh@uniyar.ac.ru} 

\date{\today} 

\begin{abstract}
We study the $\eta^\prime g^* g^*$ vertex for
arbitrary gluon virtualities in the time-like and space-like regions,
using the perturbative QCD hard scattering approach and
an input wave-function of the $\eta^\prime$-meson consistent with
the measured $\eta^\prime \gamma^* \gamma$ transition form factor.
The contribution of the gluonic content of the $\eta^\prime$-meson is
taken into account, enhancing the form factor over the entire virtuality
considered. However, data on the electromagnetic transition form factor 
of the $\eta^\prime$-meson is not sufficient to quantify the gluonic 
enhancement. We also study the effect of the transverse momenta of the
partons in the $\eta^\prime$-meson on the $\eta^\prime g^* g^*$ vertex,
using the modified hard scattering approach based on Sudakov
formalism, and contrast the results with the ones in the standard hard
scattering approach in which such effects are neglected. 
Analytic expressions for the $\eta^\prime g^* g^*$ vertex are presented in
limiting kinematic regions and parametrizations 
are given satisfying the QCD anomaly, for real gluons, and perturbative
QCD behavior for large gluon virtualities, in both the time-like and
space-like regions. Our results have implications for the
inclusive decay $B \to \eta^\prime X$ and exclusive decays, such as 
$B \to \eta^\prime (K,K^*)$, and in hadronic production processes
$N + N (\bar N) \to \eta^\prime X$.  
\end{abstract}


\maketitle

\section{\label{sec:introd}Introduction}

The vertex involving the coupling of two gluons and $\eta^\prime$-meson 
$F_{\eta^\prime g^* g^*} (q_1^2, q_2^2, m_{\eta^\prime}^2)$ (here, $q_1^2$
and $q_2^2$ represent the virtualities of the two gluons) enters in a
number of production and decay processes. For example, the  inclusive
decays $B \to \eta^\prime X_s$~\cite{cleo-etap-incl} and exclusive decays 
$B \to \eta^\prime K$~\cite{cleo-etap-excl,babar-etap}, involve, apart 
from the matrix elements of the four-quark operators, the transitions 
$b \to s g^*$, followed by $g^* \to \eta^\prime g$~\cite{TFF-3,TFF-2},
$b \to s g g $ followed by $g g \to \eta^\prime$~\cite{ACGK98}, as well 
as the transitions $g^* g^* \to \eta^\prime$ and 
$g^* g \to \eta^\prime$~\cite{TFF-1}. 
Thus, a reliable determination of the vertex function
(which can also be termed as the gluonic transition form factor) 
$F_{\eta^\prime g^* g^*} (q_1^2, q_2^2, m_{\eta^\prime}^2)$ is an
essential input in a quantitative understanding of these and related 
decays. Apart from the mentioned $B$-decays, the $\eta^\prime g^* g^{(*)}$ 
vertex plays a role in a large number of processes, among them
the radiative decay $J/\psi \to \eta^\prime \gamma$ and the hadronic
production processes $N + N (\bar N) \to \eta^\prime + X$, where~$N$ 
is a nucleon. The QCD axial anomaly~\cite{anomaly}, responsible for the 
bulk of the $\eta^\prime$-meson  mass, normalizes the vertex function on 
the gluon mass-shell, yielding $F_{\eta^\prime g^* g^*} 
(0, 0, m_{\eta^\prime}^2)$. The question that still remains concerns 
the determination of the vertex for
arbitrary time-like and space-like virtualities, $q_i^2$; $i=1, 2$.
A related aspect is to understand the relation between the
$\eta^\prime g^* g^*$ vertex and the wave-function of the
$\eta^\prime$-meson. Stated differently, issues such as the transverse
momenta of the partons in the $\eta^\prime$-meson and their impact on the
$\eta^\prime g^* g^*$ vertex have to be studied quantitatively.

  While information on the $\eta^\prime g^* g^*$ vertex is
at present both indirect and scarce, its electromagnetic counterpart
involving the coupling of two photons and the $\eta^\prime$-meson,
$F_{\eta^\prime \gamma^* \gamma}$, more generally the meson-photon
transition form factor, has been the subject of intense theoretical
and experimental activity. In particular, the hard scattering
approach to transition form factors, developed by Brodsky and Lepage
\cite{BL}, has been extensively used in studying
perturbative QCD effects and in making detailed comparison with
data~\cite{BL-Review}. 
A variation of the hard scattering approach, in which transverse degrees   
of freedom are included in the form of Sudakov effects in transition form
factors~\cite{Collins,BS89}, has also been employed in data analyses.
It has been argued~\cite{BS89,Ong,JKR96} that the Sudakov effects improve
the applicability of perturbative QCD methods down to moderate values
of~$Q^2$ and give a better account of data on meson-photon transitions
\cite{eta'-gamma}. For a critical review and comparison of the standard
(Brodsky-Lepage) and modified (mHSA) hard scattering approaches, see
Refs.~\cite{MR97,Stefanis99}. We note that either of these 
approaches combined with data
constrains the input wave-function for the quark-antiquark part of the
$\eta^\prime$-meson. However, the gluonic part of the $\eta^\prime$-meson
wave-function is not directly measured in these experiments and
will be better constrained in future experiments sensitive to
the $\eta^\prime g^* g^*$ vertex. 

A closely related issue is that of the $\eta - \eta^\prime$ mixing.
There exist good theoretical~\cite{Leutwyler} and 
phenomenological~\cite{FK98} reasons
to suspect that the frequently assumed pattern
of the decay constants $f_P^k$, defined by the equations 
\begin{equation}
\langle 0 \vert J_{\mu 5}^k \vert P(p) \rangle = i f_P^k p_\mu,
\qquad
(k = 8,1; \quad P = \eta, \eta^\prime),
\label{eq:naive-mix}
\end{equation}
where $J_{\mu 5}^8$ and $J_{\mu 5}^1$ denotes the SU(3)$_F$ octet and
singlet axial-vector currents, respectively, do not follow the pattern of
state mixing. Defining the decay constants as~\cite{Leutwyler}
\begin{eqnarray}
&&
f_{\eta}^8 = f_8 \cos \theta_8, \quad
f_{\eta}^1 = - f_1 \sin \theta_1,
\nonumber\\
&& 
f_{\eta^\prime}^8 = f_8 \sin \theta_8, \quad
f_{\eta^\prime}^1 = f_1 \cos \theta_1,
\label{eq:leutwyler-mix}
\end{eqnarray}
the angles $\theta_1$ and $\theta_8$ are found to differ considerably due
to non-negligible SU(3)$_F$-breaking effects \cite{Leutwyler,FK98}. In
contrast, it is natural
to expect that the state mixing is
determined essentially by a single
angle, as the state $\eta_c$ is far too heavy to be significant in the
state-mixing in the $\eta - \eta^\prime$ complex. A growing
consensus is now emerging in favor of
the two mixing-angle scheme of Leutwyler~\cite{Leutwyler}
for the decay constants in Eq.~(\ref{eq:leutwyler-mix})~\cite{FK98,FKS98}.
To be precise, we shall use the state-mixing scheme of Feldmann, Kroll and
Stech~\cite{FKS98}, which  is consistent with the two
mixing-angle scheme of Ref.~\cite{Leutwyler} in the current-mixing basis.

 Based on the foregoing discussion, it is natural to use the hard
scattering approach to study the $\eta^\prime g^* g^*$ vertex,
incorporating the information on the wave-function and the mixing
parameters entering in the $\eta - \eta^\prime$ complex from existing data
involving the electromagnetic transitions.
A first step in this direction was undertaken recently
by Muta and Yang~\cite{Muta}. They derived the $\eta^\prime g^* g$
vertex in the time-like region in terms of the quark-antiquark and
gluonic parts of the $\eta^\prime$-meson wave-function, taking into 
account the evolution equations obeyed by these partonic 
components~\cite{Ohrndorf:1981uz}.
In this paper, we also address the same issue along very similar lines.
We first rederive the $\eta^\prime g^* g$ vertex, pointing out
the agreement and differences between our results and the ones
in Ref.~\cite{Muta}. The latter have to do with the derivation of the
leading order perturbative contribution to the gluonic part of the 
$\eta^\prime g^* g$ vertex, and the use by Muta and Yang~\cite{Muta} of
the anomalous dimensions derived in Ref.~\cite{Ohrndorf:1981uz} in the
evolution of the wave-functions. For the anomalous dimensions, we use 
now, correcting a similar mistake in the earlier version of this paper, 
the results derived  in Refs.~\cite{Shifman:1981dk,Baier:1981pm}, which
are at variance with the ones given in Ref.~\cite{Ohrndorf:1981uz}, but
which have been  recently confirmed by Belitsky and 
M\"uller~\cite{Belitsky}. Making use of the $\eta^\prime \gamma^* \gamma$ 
data to constrain the $\eta^\prime$-meson wave-function parameters, we 
find that the gluonic contribution in the $\eta^\prime$-meson is very 
significant. We then extend our analysis to the case when
both the gluons are virtual, having either the time-like or space-like
virtualities. We also present a number of results on the vertex function
$F_{\eta^\prime g^* g^*} (q_1^2, q_2^2, m_{\eta^\prime}^2)$ in the 
asymptotic region, which
have not been presented earlier to the best of our knowledge, and are
of use in future theoretical analysis, in particular in the
$B \to \eta^\prime X$ and $B \to \eta^\prime (K, K^*)$ transitions.
We study the effects of the transverse momentum
distribution involving the constituents of the $\eta^\prime$-meson and
take into account soft-gluon emission from the constituent partons by
including the QCD Sudakov factor, following techniques {\it en vogue}
in studies of the electromagnetic and transition form factors of the 
mesons~\cite{JKR96,JK93,Stefanis99}. We show the improvements in the
$\eta^\prime g^* g^{(*)}$ vertex function due to the
inclusion of the transverse-momentum and  
Sudakov effects, which are particularly marked in the
space-like region, improving the applicability of the hard scattering
approach to low values of $Q^2$. These effects have a bearing on the 
hard scattering approach to exclusive non-leptonic decays~\cite{BBNS99}; 
the importance of the transverse-momentum and Sudakov effects in the 
decays $B \to \pi \pi$ and $B \to K\pi$ has also been recently emphasized 
in Ref.~\cite{KL00}. Finally, we also derive approximate
formulae for the $\eta^\prime g^* g^{(*)}$ vertex, which satisfy the
axial-vector anomaly result for on-shell gluons and the asymptotic
behavior in the large-$Q^2$ domain, determined by perturbative QCD.

 This paper is organized as follows: In section~\ref{sec:eta-WF}, we 
specify the $\eta - \eta^\prime$ mixing formalism and the evolution 
equations for the~$q \bar q$ and gluonic wave-functions of the 
$\eta^\prime$-meson. In section~\ref{sec:quark-contrib}, the~$q \bar q$ 
contribution to the off-shell $\eta^\prime g^* g^*$ vertex is worked out 
in the hard scattering approach. The corresponding contributions from the 
gluons are presented in section~\ref{sec:gluon-contrib}. 
In section~\ref{sec:Sudakov}, we implement the transverse-momentum effects 
and derive the Sudakov-improved $\eta^\prime g^* g^*$ vertex. Numerical 
results are presented in section~\ref{sec:numeric}, where we also show 
comparison with existing results. In section~\ref{sec:Approximate}, we 
give simple formulae for the $\eta^\prime g^* g^*$ vertex, which 
interpolate between the well-known limiting cases. Our main results are 
summarized in section~\ref{sec:concl}. Appendix~\ref{app:evol-eqns} 
contains the solution of the evolution equations for the wave functions
$\phi^{(q)} (x, Q)$ and $\phi^{(g)} (x, Q)$, and the function 
$J(\omega, \eta)$ introduced in the derivation of the 
$\eta^\prime g^* g^*$ vertex in section~\ref{sec:quark-contrib} is given 
in Appendix~\ref{app:J-func}.

\section{\label{sec:eta-WF}%
         $\eta^\prime$-Meson Wave-Function} 

The $\eta^\prime$-meson is not a flavor-octet meson state and, hence, 
in addition to the usual quark content the $\eta^\prime$-meson  
wave-function has a gluonic admixture. In principle, there also exist a 
$\bar c c$ component in the $\eta^\prime$-meson wave-function, but it has 
been estimated to be rather small in well-founded theoretical
frameworks~\cite{ACGK98,AG97,FKS98}, and hence ignored. We take the
parton Fock-state decomposition of the $\eta^\prime$-meson wave-function
as follows:  
\begin{equation} 
| \eta^\prime > \, = \sin \phi \, | \eta^\prime_q > + \cos \phi \, |
\eta^\prime_s > 
+ \, | \eta^\prime_g > ,  
\label{eq:WF-decomp}
\end{equation}
where the SU(3)$_F$ symmetry among the light $u$, $d$ and $s$ quarks 
is assumed so that $|\eta^\prime_q > \, \sim |\bar u u + \bar d d>/\sqrt 2$ 
and $| \eta^\prime_s > \, \sim | \bar s s >$ are the quark Fock states 
and~$\phi$ is the mixing angle; $|\eta^\prime_g> \, \sim |gg>$ 
is the two-gluon Fock state. 
In Refs.~\cite{Ohrndorf:1981uz,Shifman:1981dk,Baier:1981pm}
the eigenfunctions of the mixing quark~$|\bar q q>$ and gluonic~$|gg>$ 
state:  
\begin{eqnarray} 
\Psi & = & C \left [ \phi^{(q)} (x, Q) + \phi^{(g)} (x, Q) \right ] , 
\label{eq:WF-tot} \\
C & = & \sqrt 2 \, f_q \sin \phi + f_s \cos \phi,   
\label{eq:C-const}
\end{eqnarray} 
where $f_q$ and $f_s$ are the decay constants of~$| {\eta^\prime}_q >$ 
and~$|{\eta^\prime}_s>$, have been calculated by solving the evolution
equations. The result for the quark and gluonic components is presented 
as infinite series of the Gegenbauer polynomials of the indices~$3/2$ 
and $5/2$~\cite{GR} and can be found in Appendix~\ref{app:evol-eqns}. 
Previous analyses  show that it is a good approximation to consider only 
a few first terms in the expansion of the quark and gluonic wave-function. 
Here, we shall keep the leading two terms in the expansion for 
$\phi^{(q)} (x, Q)$, and keep only the first term for 
$\phi^{(g)} (x, Q)$:   
\begin{widetext}
\begin{eqnarray} 
\hspace*{-15mm} && 
\phi^{(q)} (x, Q) = 6 x \bar x  
\left \{ 1 + 
\left [ 6 B^{(q)}_2 
\left ( \frac{\alpha_s (Q^2)}{\alpha_s (\mu_0^2)} \right )^{\frac{48}{81}}
- \frac{B^{(g)}_2}{15}   
\left ( \frac{\alpha_s (Q^2)}{\alpha_s (\mu_0^2)} \right
)^{\frac{101}{81}}
\right ] (1 - 5 x \bar x) + \cdots 
\right \} , 
\label{eq:qef} \\
\hspace*{-15mm} && 
\phi^{(g)} (x, Q) = x \bar x (x -  \bar x) 
\left [ 16 B^{(q)}_2 
\left ( \frac{\alpha_s (Q^2)}{\alpha_s (\mu_0^2)} \right )^{\frac{48}{81}}
+ 5 B^{(g)}_2 
\left ( \frac{\alpha_s (Q^2)}{\alpha_s (\mu_0^2)} \right
)^{\frac{101}{81}}
\right ] + \cdots , 
\label{eq:gef} 
\end{eqnarray} 
\end{widetext} 
where $x$ and $\bar x = 1 - x$ are the energy fractions of two 
partons inside the $\eta^\prime$-meson, $Q^2 > 0$ is the energy scale 
parameter, and $\mu_0 \simeq 0.5$~GeV is the typical hadronic energy
scale below which no perturbative evolution takes place. 
%
%
\begin{figure}[b]
%
\psfig{file=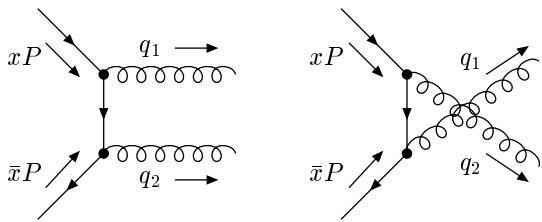,viewport=130 500 360 610,width=.45\textwidth} 
\caption{\label{fig:diag-qu}%
         Leading Feynman diagrams contributing to the quark part 
         of the $\egg$ vertex.} 
\end{figure}
%
%
It is seen that in the limit $Q^2 \to \infty$ the quark 
wave-function~(\ref{eq:qef}) turns to its asymptotic form 
$\phi_{\rm as} (x) = 6 x \bar x$ (the same asymptotic behavior as the 
pion wave-function~\cite{BL} due to its quark content), 
while the gluonic wave-function~(\ref{eq:gef}) vanishes in this limit,
$\phi^{(g)}_{\rm as} (x) = 0$.  
The coefficients of the expansion of the wave-functions~(\ref{eq:qef}) 
and~(\ref{eq:gef}) are calculated by using  perturbation theory and 
include the effective QCD coupling $\alpha_s (Q^2)$, which in the 
next-to-leading logarithmic approximation is given by~\cite{PDG}
\begin{equation} 
\alpha_s (Q^2) = \frac{4 \pi}{\beta_0 \ln (Q^2 / \Lambda^2)} 
\left [ 1 - \frac{2 \beta_1}{\beta_0^2} \, 
\frac{\ln \ln (Q^2 / \Lambda^2)}{\ln (Q^2 / \Lambda^2)} \right ],
\label{eq:alpha-s} 
\end{equation}  
where $\beta_0 = 11 - 2 n_f / 3$, $\beta_1 = 51 - 19 n_f /3$, 
$\Lambda = \Lambda_{\rm QCD}$ is the QCD scale parameter,  
and $n_f$ is the number of quarks with masses less than the energy 
scale~$Q$. In the energy region $\bar m_c < Q < \bar m_b$, where 
$\bar m_c = 1.3 \pm 0.3$~GeV and $\bar m_b = 4.3 \pm 0.2$~GeV are 
the charm and bottom $\overline{\rm MS}$ quark masses, respectively, 
we will use the value of the
dimensional parameter $\Lambda^{(4)}_{\overline{\rm MS}} = 280$~MeV
corresponding to four active quark flavors~\cite{PDG}. The requirement
that the effective QCD coupling is continuous at the flavor thresholds
leads to a change of $\Lambda$ in the threshold energy regions where
$n_f$ changes. This implies that for $Q < \bar m_c$, the value 
of~$\Lambda$ is rescaled, yielding 
$\Lambda^{(3)}_{\overline{\rm MS}} = 333$~MeV.

\section{\label{sec:quark-contrib}%
         Quark-Antiquark Contribution to the $\egg$ Vertex}

\begin{figure}[tb]
%
\psfig{file=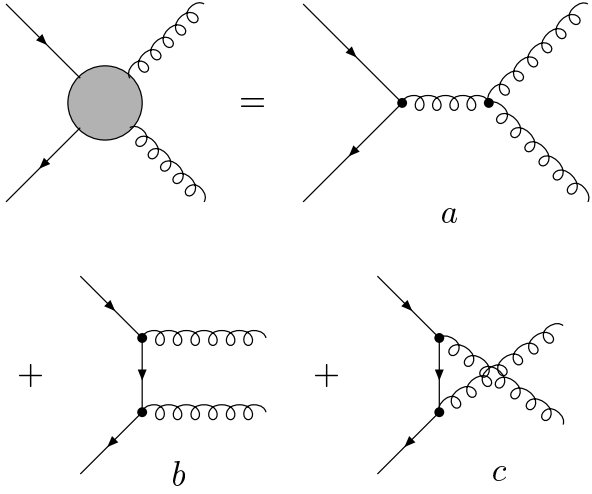,viewport=160 500 405 710,width=.45\textwidth}
\caption{\label{fig:qqgg}%
         The effective quark-antiquark-gluon-gluon vertex and 
         leading order contributions.}
\end{figure}

The diagrams depicting the quark-antiquark content of the $\egg$
vertex (or transition amplitude) are shown in Fig.~\ref{fig:diag-qu}.
The invariant amplitude corresponding to the quark 
contribution to the $\egg$ vertex in the momentum space can be defined as: 
\begin{equation} 
{\cal M}^{(q)} = \frac{C}{4 N_c} \int\limits_0^1 dx \, 
\phi^{(q)} (x, Q) \, 
{\rm Sp} [ \gamma_5 (\rlap / P - m_{\eta^\prime}) T^{(q)}_{\rm H} ] , 
\label{eq:QA-def} 
\end{equation}
where $P_\mu$ and $m_{\eta^\prime}$ are the 
four-momentum and the mass of the $\eta^\prime$-meson, respectively, 
$\rlap/ P = P_\mu \gamma^\mu$, $\gamma_\mu$ and $\gamma_5$ are
the Dirac $\gamma$-matrices, and $T^{(q)}_{\rm H}$ is the hard
amplitude connected with the effective quark-antiquark-two-gluon 
vertex by the following relation:  
\begin{equation} 
T^{(q)}_{\rm H} = \delta_{\alpha \beta} \, 
V^{\beta \alpha; a b}_{\mu \nu} (x P, \bar x P, - q_1, - q_2) \, 
\varepsilon^{a*}_\mu \varepsilon^{b*}_\nu , 
\label{eq:hard-ampl}
\end{equation}
where the summation over the quarks' colors $\alpha$ and $\beta$ allows 
to get a color-singlet meson state and $x$ and $\bar x$ have been 
defined earlier in Sec.~\ref{sec:eta-WF}.
In the leading order in the strong coupling $\alpha_s$, the effective 
quark-antiquark-gluon-gluon vertex is presented in Fig.~\ref{fig:qqgg} 
and has the form:
%
\begin{eqnarray}
&&
V^{\beta \alpha; a b}_{\mu \nu} (p_1, p_2, q_1, q_2) = 
- 4\pi \alpha_s \,
\bigg \{
\frac{i \, f_{abc} (t_c)_{\beta \alpha}}{(q_1 + q_2)^2 + i \epsilon}
\qquad 
\label{eq:qq2G-result} \\ 
&& \times 
\left [ (\rlap/ q_1 - \rlap/ q_2) g_{\mu \nu}  +
(q_1 + 2 q_2)_\mu \gamma_\nu - (2 q_1 + q_2)_\nu \gamma_\mu
\right ]
\nonumber \\
&& + 
\left ( t_b t_a \right )_{\beta \alpha}
\frac{\gamma_\nu (\rlap/ p_1 + \rlap/ q_1 + m_q) \gamma_\mu}
     {(p_1 + q_1)^2 - m_q^2 + i \epsilon} 
\nonumber \\
&& +
\left ( t_a t_b \right )_{\beta \alpha}
\frac{\gamma_\mu (\rlap/ p_1 + \rlap/ q_2 + m_q) \gamma_\nu}
     {(p_1 + q_2)^2 - m_q^2 + i \epsilon}
\bigg \} ,
\nonumber 
\end{eqnarray}
%
where $t_a$ ($a = 1, \ldots, N_c^2 - 1$) are the generators of the
color SU($N_c$) group, $f_{abc}$ are its antisymmetric
structure constants, $g_{\mu \nu} = {\rm diag} (1, - 1, -1, -1)$ is
the metric tensor, $m_q$ is the quark mass, and 
$\rlap/ q = q_\mu \gamma^\mu$.
In the calculation of the amplitude,
we have neglected all the quark masses~$m_q$ as they are much
smaller than the energy scale parameter~$Q$.
%
%
%
%

Defining the quark contribution to the $\egg$ vertex $F^{(q)}_{\egg}$ as: 
\begin{equation} 
{\cal M}^{(q)} \equiv -i \, F^{(q)}_{\egg} (q_1^2, q_2^2,
m_{\eta^\prime}^2) \, 
\delta_{a b} \, \varepsilon^{\mu \nu \rho \sigma} \, 
\varepsilon^{a*}_\mu \varepsilon^{b*}_\nu q_{1\rho} q_{2\sigma},
\label{eq:FFQ-def}
\end{equation}
the result of the calculation is: 
\begin{eqnarray} 
&& F^{(q)}_{\egg} (q_1^2, q_2^2, m_{\eta^\prime}^2) =  
4 \pi \alpha_s (Q^2) \, \frac{C}{2 N_c} 
\int\limits_0^1 dx \, \phi^{(q)} (x, Q)
\nonumber \\ 
&& \times  
\left [ \frac{1}
{x q_1^2 + \bar x q_2^2 - x \bar x m_{\eta^\prime}^2 + i \epsilon} 
+ (x \leftrightarrow \bar x) 
\right ]. 
\label{eq:QFF-result1}
\end{eqnarray}
Note that the quark wave-function of the $\eta^\prime$-meson satisfies 
the symmetry condition $\phi^{(q)} (x, Q) = \phi^{(q)} 
(\bar x,Q)$~\cite{Ohrndorf:1981uz,Shifman:1981dk,Baier:1981pm}. 
When one of the gluons is on the mass shell, for example $q_2^2 = 0$, 
the quark part of the $\egg$ vertex, $F^{(q)}_{\egg} (q_1^2, 0,
m_{\eta^\prime}^2)$, agrees with the one presented in Ref.~\cite{Muta}. 

In the case when both virtualities of the gluons are of the same signs 
(space-like $q_1^2 < 0$, $q_2^2 < 0$ or time-like $q_1^2 > 0$, $q_2^2 > 0$) 
the quark contribution to the $\egg$ vertex can be presented 
in the form:
\begin{eqnarray} 
F^{(q)}_{\egg} (q^2, \omega, \eta) & = &  
\frac{4 \pi \alpha_s (Q^2)}{q^2} \, \frac{2 C}{N_c} 
\label{eq:QFF-result2} \\ 
& \times & \int\limits_0^1 
\frac{(1 - 2 x \bar x \eta) \, \phi^{(q)} (x, Q)\, dx}
     {(1 - 2 x \bar x \eta)^2 - \omega^2 (x - \bar x)^2 + i \epsilon} , 
\nonumber 
\end{eqnarray}
where $q^2 = q_1^2 + q_2^2$ is the total gluon virtuality, 
$\omega = (q_1^2 - q_2^2)/q^2$ is the asymmetry parameter 
having the values in the domain $-1 \le \omega \le 1$, and
$\eta = m_{\eta^\prime}^2 /q^2$ is the scaled
$\eta^\prime$-meson mass squared. 
As a typical mass scale~$Q$ which defines the value of the 
strong coupling $\alpha_s (Q^2)$, it is natural to consider the 
absolute value of the total virtuality $Q^2 = |q^2|$.  
The quark contribution, $F^{(q)}_{\egg}$, corresponding 
to keeping the first two terms in the quark
wave-function~(\ref{eq:qef}) is:
%
\begin{eqnarray}
&& 
F^{(q)}_{\egg} (Q^2, \omega, \eta) =     
F^{(q)}_0 (Q^2, \omega, \eta) + F^{(q)}_2 (Q^2, \omega, \eta) 
\nonumber \\ 
&& =    
\frac{4 \pi \alpha_s (Q^2)}{m_{\eta^\prime}^2} \, \frac{6 C}{N_c} 
\left \{ {\rm G}^{(q)}_0 (\omega, \eta) + {\rm G}^{(q)}_2 (\omega, \eta) 
\vphantom{\left ( 
\frac{\alpha_s (Q^2)}{\alpha_s (\mu_0^2)} \right )^{\frac{48}{81}}}
\right. 
\label{eq:QFF-gen} \\
&& \times 
\left. 
\left [ 6 B^{(q)}_2 
\left ( \frac{\alpha_s (Q^2)}{\alpha_s (\mu_0^2)} \right )^{\frac{48}{81}}
- \frac{B^{(g)}_2}{15}   
\left ( \frac{\alpha_s (Q^2)}{\alpha_s (\mu_0^2)} \right)^{\frac{101}{81}}
\right ] 
\right \} , 
\nonumber 
\end{eqnarray}
%
where the functions ${\rm G}^{(q)}_0$ and ${\rm G}^{(q)}_2$ are: 
\begin{equation}
{\rm G}^{(q)}_0 (\omega, \eta) = - 1 +
\frac{\omega}{2 \eta} \, \ln \left | \frac{1 + \omega}{1 - \omega} \right |
+ \left [ 1 - \frac{\omega^2}{\eta} \right ] {\rm J} (\omega, \eta) ,
\label{eq:Gq0-func} 
\end{equation}
\begin{eqnarray}
{\rm G}^{(q)}_2 (\omega, \eta) & = &
- \frac{1}{6} + \frac{5}{2 \eta} - \frac{5 \omega^2}{\eta^2} 
\label{eq:Gq2-func} \\  
& + & \frac{\omega}{2 \eta} 
\left [ 1 - \frac{5}{\eta} + \frac{5 \omega^2}{\eta^2} \right ] 
\ln \left | \frac{1 + \omega}{1 - \omega} \right | 
\nonumber \\
& + & 
\left [ 1 - \frac{5 + 7 \omega^2}{2 \eta}  
+ \frac{10 \omega^2}{\eta^2} - \frac{5 \omega^4}{\eta^3}  
\right ] 
{\rm J} (\omega, \eta) . 
\nonumber 
\end{eqnarray}
Notice that ${\rm G}^{(q)}_i (\omega, \eta)$ ($i = 0, 2$) are symmetric 
in their first argument under the interchange $\omega \to -\omega$:
${\rm G}^{(q)}_i (- \omega, \eta) = {\rm G}^{(q)}_i (\omega, \eta)$. 
The function ${\rm J} (\omega, \eta)$ is defined and analyzed
in Appendix~\ref{app:J-func}. 

%
%
\begin{figure}[tb]
%
\psfig{file=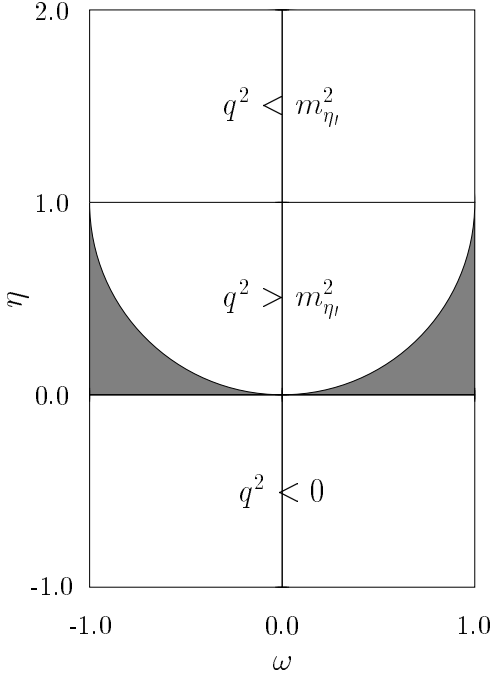,viewport=140 250 420 640,width=.40\textwidth}
\caption{\label{fig:om-et-BL}%
         The $\omega-\eta$ parameter plane of the
         $\egg$ vertex. The regions over which the vertex 
         has an imaginary part are indicated by the gray color.} 
\end{figure}
%
%

In the space-like region of the gluon virtualities, where $\eta < 0$, 
the $\egg$ vertex is real as in this case the 
quark propagator does not have poles which is clearly seen from
Eq.~(\ref{eq:QFF-result2}). In the time-like region of the virtualities, 
for those values of the asymmetry~$\omega$  and the relative
$\eta^\prime$-meson mass squared~$\eta$ for which the inequality 
$\omega^2 + (1 - \eta)^2 > 1$ is satisfied (Fig.~\ref{fig:om-et-BL}), 
an imaginary part in the vertex function appears 
due to the function $J (\omega, \eta)$ as the result of the $i \epsilon$ 
prescription of the quark propagator (see Eq.~(\ref{eq:QFF-result1})). 
Notice that if one of the gluons is on the mass shell, for example 
the second one, $q_2^2 = 0$, the imaginary part is nonzero for any value
of the other virtuality in the region $q_1^2 > m_{\eta^\prime}^2$. For
large gluon virtualities the imaginary part of the vertex
becomes large due to the linear dependence on $q^2$
of the leading contribution, ${\rm Im} \, F^{(q)}_0 \sim q^2$, and the
cubic dependence of the next-to-leading contribution, ${\rm Im} \,
F^{(q)}_2 \sim
(q^2)^3$. This rapid increase of the imaginary part of the $\egg$
vertex function with the total gluon virtuality
seems  unphysical and the natural prescription in the Brodsky-Lepage
approach is to drop the imaginary part. The physical
interpretation 
of this procedure will be discussed in Sec.~\ref{sec:Sudakov}. In the
following, we shall drop the imaginary part of the~$\egg$ vertex in 
the Brodsky-Lepage approach.

%
%
\begin{figure}[tb]
%
\psfig{file=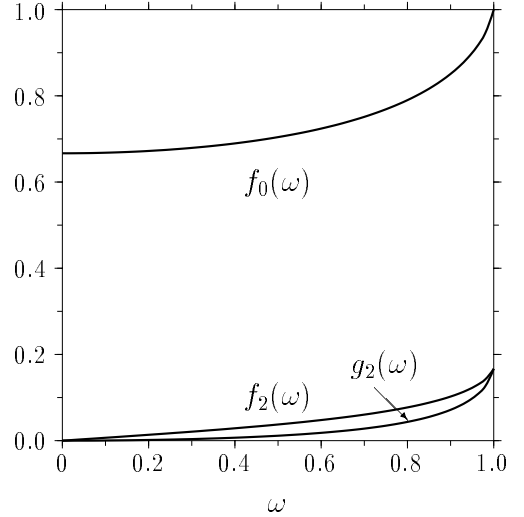,viewport=150 400 400 665,width=.40\textwidth}
\caption{\label{fig:asymm}%
         The functions $f_0 (\omega)$, $f_2 (\omega)$, and $g_2 (\omega)$ 
         describing the large $Q^2$ asymptotics of the $\egg$ 
         vertex, with the two gluon virtualities having same signs.} 
\end{figure}
%
%

It is useful to find the asymptotics of the quark contribution 
to the $\egg$ vertex at large $Q^2$ ($|\eta| \ll 1$ 
or equivalently $Q^2 \gg m_{\eta^\prime}^2$) for arbitrary
values of the asymmetry 
parameter~$\omega$. As in the case of the $\pi \gamma^* \gamma^*$ 
transition form factor the function describing the $\eta^\prime g^* g^*$
vertex decreases like $\sim 1/Q^2$: 
%
\begin{eqnarray} 
&& 
F^{(q)}_{\egg} (Q^2, \omega, 0) \simeq 
F^{(q)}_0 (Q^2, \omega, 0) + F^{(q)}_2 (Q^2, \omega, 0) 
\nonumber \\ 
&& = 
\frac{4 \pi \alpha_s (Q^2)}{Q^2} \, \frac{3 C}{N_c} \, 
\left \{ f_0 (\omega) + f_2 (\omega) 
\vphantom{\left ( 
\frac{\alpha_s (Q^2)}{\alpha_s (\mu_0^2)} \right )^{\frac{48}{81}}} 
\right.
\label{eq:QFF-1-asymp} \\
&& \times 
\left.  
\left [ 6 B^{(q)}_2 
\left ( \frac{\alpha_s (Q^2)}{\alpha_s (\mu_0^2)} \right )^{\frac{48}{81}}
- \frac{B^{(g)}_2}{15}   
\left ( \frac{\alpha_s (Q^2)}{\alpha_s (\mu_0^2)} \right )^{\frac{101}{81}}
\right ]  
\right \} , 
\nonumber 
\end{eqnarray}
%
where the two functions of the asymmetry parameter introduced above are
defined as follows: 
\begin{eqnarray} 
f_0 (\omega) & = & \frac{1}{\omega^2} 
\left [ 
1 - \frac{1 - \omega^2}{2 \omega} \ln 
\left | 
\frac{1 + \omega}{1 - \omega} 
\right |
\right ] ,
\label{eq:f-func-asymp} \\
f_2 (\omega) & = & \frac{1}{12 \omega^2} 
\left [ 3 (5 - \omega^2) f_0 (\omega) - 10
\right ]. 
\nonumber 
\end{eqnarray}
The functions $f_0(\omega)$ and $f_2(\omega)$ are displayed in 
Fig.~\ref{fig:asymm}. 
As both these functions are symmetric in their arguments: 
$f_i (- \omega) = f_i (\omega)$ ($i = 0, 2$) we present them  
in the region $0 \le \omega \le 1$ only. These functions take 
their maximum values at the extremal values of the asymmetry $\omega$: 
$f_0 (\pm 1) = 1$ and $f_2 (\pm 1) = 1/6$. At small values of 
the asymmetry parameter ($\vert \omega \vert \ll 1$), $f_2 (\omega)$
has a quadratic behavior 
$f_2 (\omega) \simeq 4 \omega^2/105$ and goes to zero with
$\omega$, while 
$f_0 (\omega)$ is finite at $\omega = 0$: $f_0 (0) = 2/3$.
This implies that if the gluon virtualities are comparable to each other,
i.e. $q_1^2 \simeq q_2^2$, the next-to-leading correction has
an additional suppression given by the function $f_2 (\omega)$.

When the virtualities of the gluons have opposite signs, 
the typical scale parameter can be defined as $Q^2 = |q_1^2 - q_2^2|$. 
In this case the quark contributions to the $\egg$ vertex is: 
%
\begin{eqnarray}
&& 
F^{(q)}_{\egg} (Q^2, \omega, \eta) = 
\frac{4 \pi \alpha_s (Q^2)}{m_{\eta^\prime}^2} \, \frac{6 C}{N_c} \, 
\frac{1}{\omega} 
\nonumber \\ 
&& \times 
\left \{
{\rm G}^{(q)}_0 \left ( \frac{1}{\omega}, \frac{\eta}{\omega} \right ) +  
{\rm G}^{(q)}_2 \left ( \frac{1}{\omega}, \frac{\eta}{\omega} \right ) 
\right.
\label{eq:QFFd-gen} \\
&& \times 
\left. 
\left [ 6 B^{(q)}_2 
\left ( \frac{\alpha_s (Q^2)}{\alpha_s (\mu_0^2)} \right )^{\frac{48}{81}}
- \frac{B^{(g)}_2}{15}   
\left ( \frac{\alpha_s (Q^2)}{\alpha_s (\mu_0^2)} \right)^{\frac{101}{81}}
\right ]
\right \} , 
\nonumber  
\end{eqnarray}
%
where in this case now $q^2 = q_1^2 - q_2^2$,
$\omega = (q_1^2 + q_2^2)/q^2$ is the asymmetry parameter,
the relative $\eta^\prime$-meson mass squared is 
$\eta = m_{\eta^\prime}^2/q^2$,
and the functions ${\rm G}^{(q)}_i$ are defined by 
Eqs.~(\ref{eq:Gq0-func}) and~(\ref{eq:Gq2-func}). 

In the region of large $Q^2$ the quark contribution to the $\egg$  
vertex has the same $1/Q^2$ behavior~(\ref{eq:QFF-1-asymp})
as in the case of the gluon virtualities of the same signs, but it has
a different  dependence on the asymmetry parameter
$f_i (\omega) \to \tilde f_i (\omega)$ ($i = 0,2$)
where: 
\begin{eqnarray}
\tilde f_0 (\omega) & = & \frac{1}{\omega^2} \,
f_0 \left ( \frac{1}{\omega} \right ) =
2 - \omega^2 f_0 (\omega) ,
\label{eq:f-func-asymp-pm} \\
\tilde f_2 (\omega) & = & \frac{1}{\omega^2} \,
f_2 \left ( \frac{1}{\omega} \right ) =
- \frac{4}{3} + \frac{5 \omega^2}{2}  +
\frac{\omega^2}{4}  (1 - 5 \omega^2) f_0 (\omega) .
\nonumber
\end{eqnarray}
The function $f_0 (\omega)$ is defined in
Eq.~(\ref{eq:f-func-asymp}), and
the curves corresponding to these functions are presented in
Fig.~\ref{fig:asym-pm}.
Note that these functions are also symmetric in the asymmetry
parameter:  $\tilde f_i (- \omega) = \tilde f_i (\omega)$, like the
functions $f_i (\omega)$ encountered earlier.

If one of the gluons is on the mass shell, for example the second 
one ($q_2^2 = 0$), the leading and next-to-leading contributions 
are simplified:   
\begin{widetext}
\begin{eqnarray} 
F^{(q)}_0 (q_1^2, 0, m_{\eta^\prime}^2) & = &
- \frac{4 \pi \alpha_s (|q_1^2|)}{m_{\eta^\prime}^2} \, \frac{6 C}{N_c} 
\left \{  
1 + \frac{q_1^2}{m_{\eta^\prime}^2} 
\ln \left ( 1 - \frac{m_{\eta^\prime}^2}{q_1^2} \right ) 
\right \} , 
\label{eq:QFF-0-shell} \\
F^{(q)}_2 (q_1^2, 0, m_{\eta^\prime}^2) & = & 
- \frac{4 \pi \alpha_s (|q_1^2|)}{m_{\eta^\prime}^2} \, \frac{6 C}{N_c} \, 
\left [ 6 B^{(q)}_2 
\left ( \frac{\alpha_s (|q_1^2|)}{\alpha_s (\mu_0^2)} \right
)^{\frac{48}{81}}
- \frac{B^{(g)}_2}{15}   
\left ( \frac{\alpha_s (|q_1^2|)}{\alpha_s (\mu_0^2)} \right)^{\frac{101}{81}}
\right ] 
\label{eq:QFF-1-shell} \\
& \times & 
\left \{ 
\frac{1}{6} - \frac{5 q_1^2}{2 m_{\eta^\prime}^2} + \frac{5
(q_1^2)^2}{m_{\eta^\prime}^4}
+ \frac{q_1^2}{m_{\eta^\prime}^2} \left [ 
1 - \frac{5 q_1^2}{m_{\eta^\prime}^2} + \frac{5
(q_1^2)^2}{m_{\eta^\prime}^4}
\right ]  
\ln \left ( 1 - \frac{m_{\eta^\prime}^2}{q_1^2} \right ) 
\right \} ~. 
\nonumber 
\end{eqnarray}
In the limit of the large gluon virtuality
($|q_1^2| \gg m_{\eta^\prime}^2$) 
the quark contribution to the $\egg$ vertex has the form: 
\begin{equation} 
F^{(q)}_{\egg} (q_1^2, 0, 0) = 
\frac{4 \pi \alpha_s (|q_1^2|)}{q_1^2} \, \frac{3 C}{N_c} \, 
\left [ 1 + B^{(q)}_2 
\left ( 
\frac{\alpha_s (|q_1^2|)}{\alpha_s (\mu_0^2)} 
\right )^{\frac{48}{81}}
- \frac{B^{(g)}_2}{90}   
\left ( 
\frac{\alpha_s (|q_1^2|)}{\alpha_s (\mu_0^2)} 
\right )^{\frac{101}{81}}
\right ] . 
\label{eq:QFF-asympt-large}
\end{equation}
When the gluon virtuality is time-like $q_1^2 > 0$,  
the quark contribution has a logarithmic 
divergence near the threshold $q_1^2=m_{\eta^\prime}^2$: 
\begin{eqnarray} 
F^{(q)}_{\egg} \bigg |_{q_1^2 \simeq m_{\eta^\prime}^2} & \simeq &   
- \frac{4 \pi \alpha_s (m_{\eta^\prime}^2)}{m_{\eta^\prime}^2} \, 
\frac{6 C}{N_c} \, 
\left \{ 
1 + \ln \left ( 1 - \frac{m_{\eta^\prime}^2}{q_1^2} \right ) 
\right.
\label{eq:QFF-asympt-thresh} \\ 
&& +  
\left. 
\left [ 6 B^{(q)}_2 
\left ( \frac{\alpha_s (m_{\eta^\prime}^2)}{\alpha_s (\mu_0^2)} \right
)^{\frac{48}{81}}
- \frac{B^{(g)}_2}{15}   
\left ( \frac{\alpha_s (m_{\eta^\prime}^2)}{\alpha_s (\mu_0^2)} \right
)^{\frac{101}{81}}
\right ] 
\left [ 
\ln \left ( 1 - \frac{m_{\eta^\prime}^2}{q_1^2} \right ) + \frac{8}{3} 
\right ]  
\right \} . 
\nonumber 
\end{eqnarray}
\end{widetext}
For the space-like gluon virtuality $q_1^2 < 0$ it is worth noting 
that Eqs.~(\ref{eq:QFF-0-shell}) and~(\ref{eq:QFF-1-shell}) are 
suitable when the gluon four-momentum satisfies the condition: 
%
%
\begin{figure}[tb]
%
\psfig{file=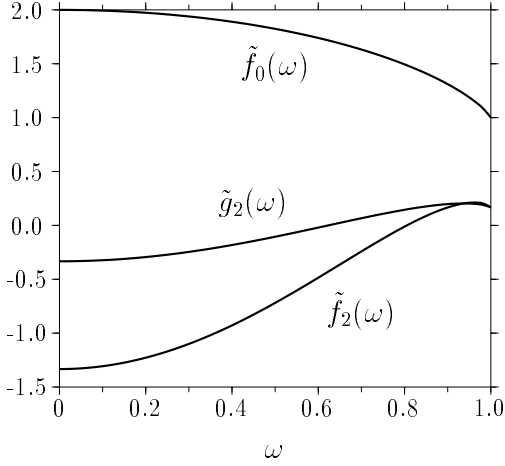,viewport=150 430 400 660,width=.40\textwidth}
\caption{\label{fig:asym-pm}%
         The functions $\tilde f_0 (\omega)$, $\tilde f_2 (\omega)$,
         and $\tilde g_2 (\omega)$ describing the asymptotics of
         the $\egg$ vertex function at large $Q^2$, with the
         two gluon virtualities having opposite signs.}
\end{figure}
%
%
$|q_1^2| > m_{\eta^\prime}^2$. In the case of the smaller absolute values 
of the gluon virtuality the $\eta^\prime$-meson mass, $m_{\eta^\prime}$,
becomes the largest scale parameter. The strong coupling $\alpha_s$
should be estimated at that value of the scale parameter which corresponds
to the following change in Eqs.~(\ref{eq:QFF-0-shell})
and~(\ref{eq:QFF-1-shell}): 
$\alpha_s (Q^2) \to \alpha_s (m_{\eta^\prime}^2)$.

\section{\label{sec:gluon-contrib}%
         Gluon Contribution to the $\egg$ Vertex}

\begin{figure}[tb]
%
\psfig{file=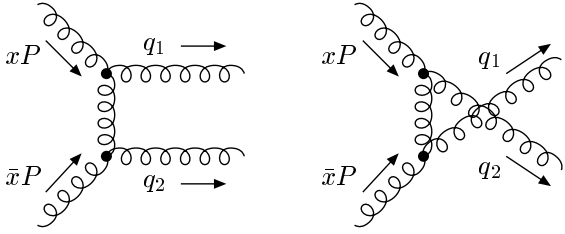,viewport=140 500 360 610,width=.45\textwidth}
\caption{\label{fig:diag-gl}%
         Leading order contribution to the gluonic part 
         of the $\egg$ vertex.}
\end{figure}

\begin{figure}[tb]
%
\psfig{file=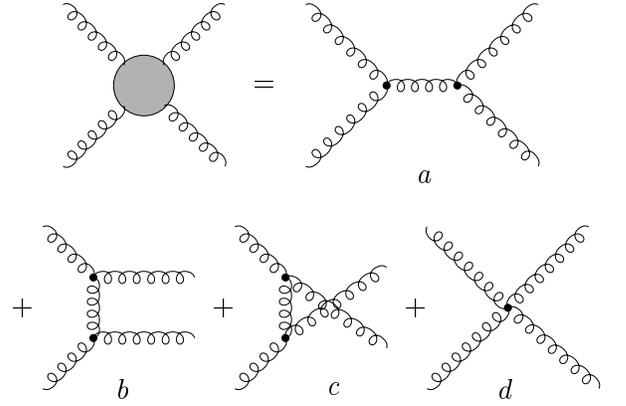,viewport=140 500 440 710,width=.45\textwidth}
\caption{\label{fig:4g}%
         The effective four-gluon vertex and leading order contributions.}
\end{figure}

The invariant amplitude of the gluon contribution to the $\egg$
vertex can be defined as: 
\begin{equation} 
{\cal M}^{(g)} = - \frac{i C}{2 N_c Q^2}
\int\limits_0^1 dx \, \phi^{(g)} (x, Q) \, 
\varepsilon^{\alpha \beta \mu \nu} q_{1\alpha} q_{2\beta} 
\left [ T^{(g)}_{\rm H} \right ]_{\mu \nu} , 
\label{eq:GA-def} 
\end{equation}
where $T^{(g)}_{\rm H}$ is the hard amplitude of the gluonic
content shown in Fig.~\ref{fig:diag-gl}.
%
%
%
%
In the Brodsky-Lepage approach it is connected with the effective
four-gluon vertex by the following relation:  
\begin{equation} 
\left [ T^{(g)}_{\rm H} \right ]_{\rho \sigma} = \delta_{c d} \, 
V^{a b c d}_{\mu \nu \rho \sigma} (- q_1, -q_2, x P, \bar x P) \, 
\varepsilon^{a*}_\mu \varepsilon^{b*}_\nu , 
\label{eq:G-hard-ampl}
\end{equation}
where the summation over the gluons' colors $c$ and $d$ allows 
to get the two-gluon color-singlet state.
In the leading order in the strong coupling $\alpha_s$ the effective
four-gluon vertex has the contributions from the diagrams of the four-gluon
annihilation shown in Fig.~\ref{fig:4g} and can be written in the form:
%
\begin{eqnarray}
\hspace{-5mm} 
&&
V^{a b c d}_{\alpha \beta \gamma \delta} (q_1, q_2, q_3, q_4) = 
- \frac{4\pi \alpha_s \, f_{abn} f_{cdn}}{(q_1 + q_2)^2 + i \epsilon}
\big \{ (q_1 + q_2)^2 
\label{eq:4G-result} \\
\hspace{-5mm}
&& \times  
[ g_{\alpha \gamma} g_{\beta \delta} - g_{\alpha \delta} g_{\beta \gamma}
]
+ g_{\alpha \beta} g_{\gamma \delta} \left ( (q_1 - q_2) (q_3 -
q_4) \right )
\nonumber \\
\hspace{-5mm}
&& + 
g_{\alpha \beta} \left [ (q_3 + 2 q_4)_\gamma (q_1 - q_2)_\delta
- (2 q_3 + q_4)_\delta (q_1 - q_2)_\gamma \right ]
\nonumber \\
\hspace{-5mm}
&& +  
g_{\gamma \delta} \left [ (q_1 + 2 q_2)_\alpha (q_3 - q_4)_\beta
- (2 q_1 + q_2)_\beta (q_3 - q_4)_\alpha \right ]
\nonumber \\
\hspace{-5mm}
&& + 
g_{\alpha \gamma} (2 q_1 + q_2)_\beta (2 q_3 + q_4)_\delta
+ g_{\beta \delta} (q_1 + 2 q_2)_\alpha (q_3 + 2 q_4)_\gamma
\nonumber \\
\hspace{-5mm}
&& -  
g_{\alpha \delta} (2 q_1 + q_2)_\beta (q_3 + 2 q_4)_\gamma
- g_{\beta \gamma} (q_1 + 2 q_2)_\alpha (2 q_3 + q_4)_\delta
\big \}
\nonumber \\
\hspace{-5mm}
&& +  
\left \{ (a, \alpha, q_1) \leftrightarrow (c, \gamma, q_3) \right \}
+ \left \{ (a, \alpha, q_1) \leftrightarrow (d, \delta, q_4) \right \} .  
\nonumber
\end{eqnarray}
%
%
%
%
Let us define the gluonic contribution, $F^{(g)}_{\egg}$, 
to the $\egg$ vertex as: 
\begin{equation} 
{\cal M}^{(g)} \equiv - i \, F^{(g)}_{\egg} \, \delta_{a b} \, 
\varepsilon^{\mu \nu \rho \sigma} \, 
\varepsilon^{a*}_\mu \varepsilon^{b*}_\nu q_{1\rho} q_{2\sigma} .
\label{eq:FFG-def}
\end{equation}
The result of the calculation is: 
\begin{eqnarray} 
&& 
F^{(g)}_{\egg} (q_1^2, q_2^2, m_{\eta^\prime}^2) = 
\frac{4 \pi \alpha_s (Q^2)}{Q^2} \, \frac{C}{2} \int\limits_0^1 
dx \, \phi^{(g)} (x, Q) 
\nonumber \\
&& \times 
\left [
\frac{x q_1^2 + \bar x q_2^2 - (1 + x \bar x) m_{\eta^\prime}^2}
     {\bar x q_1^2 + x q_2^2 - x \bar x m_{\eta^\prime}^2 + i \epsilon} 
- (x \leftrightarrow \bar x) \right ] . 
\label{eq:GFF-result1}
\end{eqnarray}
Note that the gluonic wave-function of the $\eta^\prime$-meson
satisfies the antisymmetry condition 
$\phi^{(g)} (x, Q) = - \phi^{(g)} (\bar x,
Q)$~\cite{Ohrndorf:1981uz,Shifman:1981dk,Baier:1981pm}. 
It implies that if the relative sign in the brackets {\it were} a ``$+$'' 
(as given in Eq.~(6) in Ref.~\cite{Muta}) the gluonic contribution 
to the $\egg$ vertex would vanish identically. We disagree with
the result stated in Ref.~\cite{Muta} on this point.
 
When the gluon virtualities are of the same signs, the gluonic 
contribution to the $\egg$ vertex can be presented in the form:
\begin{eqnarray} 
&& 
F^{(g)}_{\egg} (Q^2, \omega, \eta) = 
\frac{4 \pi \alpha_s (Q^2)}{Q^2} \, 2 C \omega 
\label{eq:GFF-result2} \\
&& \times 
\int\limits_0^1 
\frac{(x - \bar x) \, [1 - (1 + 2 x \bar x) \eta] \, 
      \phi^{(g)} (x, Q) \,dx}
     {(1 - 2 x \bar x \eta)^2 - \omega^2 (x - \bar x)^2 + i \epsilon} ,  
\nonumber
\end{eqnarray}
where the total gluon virtuality~$q^2$, the asymmetry parameter~$\omega$, 
and~$\eta$ are the same as the ones entering in
Eq.~(\ref{eq:QFF-result2}) and we have set the
scale $Q^2 = |q^2|$. After the integration, and taking into account 
the gluonic wave-function~(\ref{eq:gef}), the result is: 
\begin{eqnarray} 
&& 
F^{(g)}_{\egg} (Q^2, \omega, \eta) = 
\frac{4 \pi \alpha_s (Q^2)}{m_{\eta^\prime}^2} \, \frac{C q^2}{Q^2} \, 
{\rm G}^{(g)}_2 (\omega, \eta) 
\qquad 
\label{eq:GFF-result3} \\
&& \times 
\left [ 16 B^{(q)}_2 
\left ( \frac{\alpha_s (Q^2)}{\alpha_s (\mu_0^2)} \right )^{\frac{48}{81}}
\! \! \! 
+ 5 B^{(g)}_2 
\left ( \frac{\alpha_s (Q^2)}{\alpha_s (\mu_0^2)} \right )^{\frac{101}{81}}
\right ], 
\nonumber 
\end{eqnarray}
where the function ${\rm G}^{(g)}_2$ has the following expression: 
\begin{eqnarray}
\hspace{-1mm} 
{\rm G}^{(g)}_2 (\omega, \eta) & = & \omega \Bigg \{ 
\frac{5}{3} + \frac{2}{\eta} - \frac{4 \omega^2}{\eta^2} 
\label{eq:Gg1-func} \\  
\hspace{-1mm} 
& + & \frac{1}{2 \omega} 
\left [ 1 - \frac{\omega^2}{\eta} \right ] 
\left [ 1 - \frac{4 \omega^2}{\eta^2} \right ] 
\ln \left | \frac{1 + \omega}{1 - \omega} \right | 
\nonumber \\
\hspace{-1mm} 
& + & 
\eta \, \left [ 1 - \frac{2}{\eta} - \frac{2 + \omega^2}{\eta^2} 
+ \frac{8 \omega^2}{\eta^3} - \frac{4 \omega^4}{\eta^4}  
\right ] 
{\rm J} (\omega, \eta) 
\Bigg \} .
\nonumber 
\end{eqnarray}
Note that this function is antisymmetric in its first argument
under the change $\omega \to -\omega$: 
${\rm G}^{(g)}_2 (- \omega, \eta) = - {\rm G}^{(g)}_2 (\omega, \eta)$. 

As in the case of the quark-antiquark contribution to the $\egg$ vertex,
the gluonic contribution also contains 
an imaginary part due to the function ${\rm J} (\omega, \eta)$ as a 
result of the $i \epsilon$ prescription of the gluon propagator. The 
imaginary part of the gluonic contribution has a cubic dependence on 
the total gluon virtuality, ${\rm Im} \, F^{(g)}_{\eta^\prime g^* g^*} 
\sim (q^2)^3$, in complete analogy with the next-to-leading quark 
contribution. We use the Brodsky-Lepage prescription discussed in the 
preceding section and drop the imaginary part in the analysis of the 
vertex function.   

The gluonic contribution to the $\egg$ vertex has the 
same $1/Q^2$ asymptotics at large $Q^2$ ($\eta \to 0$) as the quark 
contribution~(\ref{eq:QFF-1-asymp}): 
\begin{eqnarray} 
&& 
F^{(g)}_{\egg} (Q^2, \omega, 0) = 
\frac{4 \pi \alpha_s (Q^2)}{Q^2} \, C \, g_2 (\omega)  
\label{eq:GFF-asymp} \\
&& \times 
\left [ 16 B^{(q)}_2 
\left ( \frac{\alpha_s (Q^2)}{\alpha_s (\mu_0^2)} \right )^{\frac{48}{81}}
\! \! \! 
+ 5 B^{(g)}_2 
\left ( \frac{\alpha_s (Q^2)}{\alpha_s (\mu_0^2)} \right
)^{\frac{101}{81}}
\right ] , 
\nonumber \\ 
&& 
g_2 (\omega) = \frac{3 f_0 (\omega) - 2}{6\omega} , 
\label{eq:g-func-asymp} 
\end{eqnarray}
where the function $f_0 (\omega)$ is defined by Eq.~(\ref{eq:f-func-asymp}). 
Note that the gluonic function $g_2 (\omega)$ is antisymmetric: 
$g_2 (- \omega) = - g_2 (\omega)$, in contrast with the quark functions 
$f_i (-\omega)=f_i (\omega)$ ($i = 0,2$). The curve corresponding to $g_2
(\omega)$ is 
presented in Fig.~\ref{fig:asymm}. It takes its maximum and minimum 
values at the borders of the argument domain: $g_2 (\pm 1) = \pm 1/6$ and 
$g_2 (0) = 0$, respectively, because of the antisymmetry condition. Again,
as in the next-to-leading order quark contribution 
to the $\egg$ vertex, the gluonic contribution 
has an additional suppression due to the function 
$g_2 (\omega)$ for the gluons with comparable virtualities. 

For the case when the gluons virtualities have opposite signs 
($q_1^2 > 0$, $q_2^2 < 0$ or $q_1^2 < 0$, $q_2^2 > 0$), the $\egg$ 
vertex can be presented in the form 
given in Eq.~(\ref{eq:GFF-result3})  but with a different dependence on
the parameters~$\omega$ and~$\eta$: 
\begin{equation}
{\rm G}^{(g)}_2 (\omega, \eta) \to 
{\rm G}^{(g)}_2 \left ( \frac{1}{\omega}, \frac{\eta}{\omega} \right ) .
\label{eq:Gg1-func-pm} 
\end{equation}
The large $Q^2$ behavior of the $\egg$ vertex is the
same as in the case of the gluon virtualities of the same signs
[Eq.~(\ref{eq:GFF-asymp})] but involves a different function of the
asymmetry parameter:
\begin{equation}
g_2 (\omega) \to \tilde g_2 (\omega) = \frac{1}{\omega} \,
g_2 \left ( \frac{1}{\omega} \right )
= - \frac{1}{3} + \omega^2 - \frac{\omega^4}{2} f_0 (\omega) .
\label{eq:g-func-asymp-pm}
\end{equation}
The function~$\tilde g_2 (\omega)$ is shown graphically in
Fig.~\ref{fig:asym-pm}.
In contrast to the function $g_2 (\omega)$, given
in Eq.~(\ref{eq:g-func-asymp}),
which is antisymmetric in its argument, the function $\tilde g_2 (\omega)$
is symmetric: $\tilde g_2 (- \omega) = \tilde g_2 (\omega)$.

When one of the gluons is on the mass shell, say, the second
gluon ($q_2^2 = 0$), the leading gluonic contribution to the form
factor is:  
\begin{widetext}
\begin{eqnarray} 
F^{(g)}_{\egg} (q_1^2, 0, m_{\eta^\prime}^2) & = & 
\frac{4 \pi \alpha_s (|q_1^2|)}{m_{\eta^\prime}^2} \, \frac{C
q_1^2}{|q_1^2|}  
\left [ 16 B^{(q)}_2 
\left ( \frac{\alpha_s (|q_1^2|)}{\alpha_s (\mu_0^2)} \right)^{\frac{48}{81}}
\! \! + 5 B^{(g)}_2   
\left ( \frac{\alpha_s (|q_1^2|)}{\alpha_s (\mu_0^2)} \right
)^{\frac{101}{81}}
\right ] 
\label{eq:GFF-1-shell} \\
& \times & 
\left \{ 
\frac{5}{3} + \frac{2 \, q_1^2}{m_{\eta^\prime}^2} - \frac{4
(q_1^2)^2}{m_{\eta^\prime}^4}
- \left [ 1 - \frac{q_1^2}{m_{\eta^\prime}^2} \right ] 
    \left [ 1 - \frac{4 (q_1^2)^2}{m_{\eta^\prime}^4} \right ]  
\ln \left ( 1 - \frac{m_{\eta^\prime}^2}{q_1^2} \right ) 
\right \} ~. 
\nonumber 
\end{eqnarray}
\end{widetext}
In the limit of the large gluon virtuality 
($|q_1^2| \gg m_{\eta^\prime}^2$) 
the gluon contribution to the $\egg$ vertex takes the form: 
\begin{eqnarray} 
&& 
F^{(g)}_{\egg} (q_1^2, 0, 0) = 
\frac{4 \pi \alpha_s (|q_1^2|)}{|q_1^2|} \, \frac{C}{6} \, 
\label{eq:GFF-asympt-large} \\
&& \times 
\left [ 16 B^{(q)}_2 
\left ( \frac{\alpha_s (|q_1^2|)}{\alpha_s (\mu_0^2)} \right
)^{\frac{48}{81}}
+ 5 B^{(g)}_2   
\left ( \frac{\alpha_s (|q_1^2|)}{\alpha_s (\mu_0^2)} \right
)^{\frac{101}{81}}
\right ] . 
\nonumber
\end{eqnarray}
As noted in the case of the quark contribution, when the gluon virtuality
is time-like, $q_1^2 > 0$, there is a threshold near $q_1^2
=m_{\eta^\prime}^2$, but
unlike the quark case, the gluonic contribution is regular, 
\begin{eqnarray} 
&& 
F^{(g)}_{\egg} (m_{\eta^\prime}^2, 0, m_{\eta^\prime}^2) = 
\frac{4 \pi \alpha_s (m_{\eta^\prime}^2)}{m_{\eta^\prime}^2} \,
\frac{C}{3} \, 
\label{eq:GFF-asympt-thresh} \\
&& \times 
\left [ 16 B^{(q)}_2 
\left ( \frac{\alpha_s (m_{\eta^\prime}^2)}{\alpha_s (\mu_0^2)} \right
)^{\frac{48}{81}}
+ 5 B^{(g)}_2  
\left ( \frac{\alpha_s (m_{\eta^\prime}^2)}{\alpha_s (\mu_0^2)} \right
)^{\frac{101}{81}}
\right ] . 
\nonumber
\end{eqnarray}
We point out that
Eq.~(\ref{eq:GFF-1-shell}) is to be used for $|q_1^2| >
m_{\eta^\prime}^2$ for the space-like gluon virtuality.
 In the region
$0 < |q_1^2| < m_{\eta^\prime}^2$ the $\eta^\prime$-meson mass is the 
largest scale parameter and in Eq.~(\ref{eq:GFF-1-shell}) the following 
change should be done: $\alpha_s (Q^2) \to \alpha_s (m_{\eta^\prime}^2)$.

\section{\label{sec:Sudakov}%
         Sudakov-Improved $\eta^\prime g^* g^*$ Vertex}

In discussing the $\eta^\prime g^* g^*$ vertex for the situation 
in which one of the gluons or both are far from the mass shell, 
the typical mass scale~$Q^2$ can be defined by the largest virtuality
$q_1^2$ or $q_2^2$ ($|q_i^2| \gg m_{\eta^\prime}^2$).
If the transverse momenta ${\bf k}_{\perp i}$ of the
partons in the $\eta^\prime$-meson
are taken into account in the hard scattering approach, the
perturbative expansion of the vertex encounters large
logarithms of the form $\ln (Q^2 / {\bf k}_{\perp i}^2)$, and it becomes
mandatory to sum the  multiple-gluon emissions. The formalism for the 
soft and collinear gluon resummation was introduced by Collins and
Soper~\cite{CS} and Collins, Soper and Sterman~\cite{CSS}. Such gluon
emissions give rise to powers of double logarithms in each order of
perturbation theory and their contribution exponentiates into the 
Sudakov function~\cite{Sudakov}. The Sudakov exponents are known both
for quarks from the Drell-Yan (DY) process and the deep inelastic
scattering (DIS), and for gluons from the gluon fusion into $2 \gamma$,
gauge, or Higgs bosons final states. This formalism is also suitable 
for the description of the hadronic wave-functions and the hadronic
form factors, such as the electromagnetic and transition form factors 
of the pion~\cite{BS89,JKR96,JK93,Sudakov-gen,Gousset}, and is also of 
interest here.

In the modified Hard Scattering Approach (mHSA)~\cite{BS89}, the
quark and gluonic invariant amplitudes which define the
$\egg$ vertex can be written as:
\begin{eqnarray}
{\cal M}^{(q)} & = & \frac{1}{\sqrt{2 N_c}} \int\limits_0^1 dx
\int \frac{d{\bf b}}{4 \pi} \,
\hat \Psi^{(q)} (x, Q, {\bf b}) \, S^{(q)} (x, Q, b)
\nonumber \\
& \times &
\int \frac{d {\bf k}_\perp}{(2 \pi)^2} \,
{\rm e}^{- i ({\bf k}_\perp {\bf b})} \,
{\rm Sp} \{ \gamma_5 ( \, \rlap / \,\!\! P - m_{\eta'}) T^{(q)}_H \} ,
\qquad
\label{eq:FFq-mHSA} \\
{\cal M}^{(g)} & = & \frac{- 2 i}{\sqrt{2 N_c}} \int\limits_0^1 dx
\int \frac{d{\bf b}}{4 \pi} \,
\hat \Psi^{(g)} (x, Q, {\bf b}) \, S^{(g)} (x, Q, b)
\nonumber \\ 
& \times &
\int \frac{d {\bf k}_\perp}{(2 \pi)^2} \,
{\rm e}^{- i ({\bf k}_\perp {\bf b})} \,
\varepsilon^{\alpha \beta \mu \nu} \,
\frac{q_{1\alpha} q_{2\beta}}{Q^2} \,
[ T^{(g)}_H ]_{\mu \nu} ,
\qquad 
\label{eq:FFg-mHSA} 
\end{eqnarray}
where ${\bf b}$ is the separation between the $\eta^\prime$ constituents
in the transverse configuration space, often called the
impact parameter, ${\bf k}_\perp$ is the transverse momentum of one 
of the constituents in the $\eta^\prime$-meson rest-frame,
and the functions $T^{(q)}_H$ and $T^{(g)}_H$ are defined in
Eqs.~(\ref{eq:hard-ampl}) and~(\ref{eq:G-hard-ampl}), respectively.
In this approach we take the $\eta^\prime$-meson wave-function in the 
form similar to the pion wave-function~\cite{JK93,JKR96}:
\begin{equation}
\hat \Psi^{(p)} (x, Q, {\bf b}) =
\frac{2 \pi C}{\sqrt{2 N_c}} \,
\phi^{(p)} (x, Q) \,
\exp \left [ - \frac{x \bar x b^2}{4 a^2} \right ] ,
\label{eq:WF-full}
\end{equation}
where $p = q$ for quark or antiquark and $p = g$ for gluons, 
the constant~$C$ is defined in Eq.~(\ref{eq:C-const}), 
and $\phi^{(q)}$ and $\phi^{(g)}$ have the forms presented in 
Eqs.~(\ref{eq:qef}) and~(\ref{eq:gef}), respectively. 
We assume that the transverse parts of the quark and gluonic 
components of the $\eta^\prime$-meson wave-function are universal 
and have a simple Gaussian distribution in the separation length
$b = |{\bf b}|$, i.e. $\hat \Psi^{(p)} (x, Q, {\bf b}) = \hat \Psi^{(p)}
(x, Q, b)$. The parameter $a$ can be determined from the average 
transverse momentum of the $\eta^\prime$-meson.
For the numerical analysis we take the transverse size
parameter $a^{-1} = 0.861$~GeV, obtained for the $\pi$-meson by Kroll 
et al.~\cite{JKR96,JK93}. The soft-gluon emission from the quark,
antiquark and gluons inside the $\eta^\prime$-meson can be taken into
account by including the QCD Sudakov factors $S^{(p)} (x, Q, b)$
in Eqs.~(\ref{eq:FFq-mHSA}) and~(\ref{eq:FFg-mHSA}).

We shall perform our analysis by taking into account the next-to-leading
logarithmic (NLL) contribution to the Sudakov factors, which are known
for both the quark~\cite{SF-quark} and gluonic~\cite{SF-gluon,Balazs}
cases.
In applications to form factors of mesons the Sudakov factor
is~\cite{Sudakov-gen}:
\begin{widetext}
\begin{equation}
S^{(p)} (x, Q, b) =
\exp \left [ - s_p (x Q, b, C_1, C_2) - s_p (\bar x Q, b, C_1, C_2)
- \int\limits_{C_1^2/b^2}^{t^2} \frac{d \mu^2}{\mu^2}
\gamma_p \left ( \alpha_s (\mu^2) \right )
\right ],
\label{eq:S-meson}
\end{equation}
where the Sudakov function $s_p (Q, b, C_1, C_2)$ for an external
quark ($p = q$) or gluon ($p = g$) line can be presented in the
following general form~\cite{CSS}:
\begin{equation}
s_p (Q, b, C_1, C_2) = \exp
\left \{ - \frac{1}{4}
\int\limits^{C_2^2 Q^2}_{C_1^2 / b^2}\frac{dq^2}{q^2}
\left [
A_p (\alpha_s (q^2)) \ln \frac{C_2^2 Q^2}{q^2} + B_p (\alpha_s (q^2))
\right ]
\right \} .
\label{eq:S-gen}
\end{equation}
\end{widetext}
The integral term in Eq.~(\ref{eq:S-meson}) arises from the application
of the renormalization group to the $\eta^\prime$-meson wave-function as
well as to the hard scattering amplitude. The upper limit $t^2$ is defined
by the largest mass scale appearing in the hard scattering amplitude.
The functions $\gamma_p (\alpha_s)$ are the anomalous dimensions of the
partons inside the $\eta^\prime$-meson in the axial gauge~\cite{BDS}:
\begin{equation}
\gamma_q (\alpha_s) = - \frac{3 C_F \alpha_s}{4 \pi} 
+ {\cal O} (\alpha_s^2),
\label{eq:anom-dim}  
\quad
\gamma_g (\alpha_s) = - \frac{3 \beta_0 \alpha_s}{4 \pi}  
+ {\cal O} (\alpha_s^2).
\end{equation}
Notice that as the energy parameter of the Sudakov function one should
use the energies~$x Q$ and~$\bar x Q$ of the $\eta^\prime$-meson
constituents which are, in general, different.
The functions $A_p (\alpha_s)$ and $B_p (\alpha_s)$ are perturbatively
computable as power series in the strong coupling constant~$\alpha_s$:
\begin{equation}
A_p (\alpha_s) = \sum_{n = 1}^\infty
\left ( \frac{\alpha_s}{\pi} \right )^n \!\! A_p^{(n)} ,
\label{eq:AB-expansion} 
\quad 
B_p (\alpha_s) = \sum_{n = 1}^\infty
\left ( \frac{\alpha_s}{\pi} \right )^n \!\! B_p^{(n)} ,
\end{equation}
where $A_p^{(n)}$ and $B_p^{(n)}$ are the expansion coefficients,
specified below for~$n = 1$ and~$n = 2$. 
We recall that the coefficients $A_p^{(1)}$ are universal and lead to 
the resummation of the leading logarithmic (LL) contributions.
The coefficients $A_p^{(2)}$ and $B_p^{(1)}$ give the NLL terms.
These coefficients
are also process-independent but they depend on the renormalization and
factorization schemes through the parameters~$C_1$ and~$C_2$.
The next expansion coefficients are process-dependent,
as demonstrated recently for~$B_p^{(2)}$~\cite{CdFG}. 
However, to the NLL accuracy, we do not require them.
The coefficients of the LL and NLL terms of the Sudakov exponent
for quarks~\cite{CSS,SF-quark,DS} and gluons~\cite{SF-gluon,Balazs} are:
\begin{eqnarray}
A_q^{(1)} & = & C_F , \qquad  A_g^{(1)} = C_A ,
\nonumber \\
A_q^{(2)} & = & \frac{1}{2} C_F \left [ K +
     \beta_0 \ln \frac{C_1}{b_0} \right ] , 
\nonumber \\ 
A_g^{(2)} & = & \frac{1}{2} C_A \left [ K +
     \beta_0 \ln \frac{C_1}{b_0} \right ] ,
\label{eq:coeff} \\
B_q^{(1)} & = & - \frac{1}{2} C_F \left [ 1 +
      2 \ln \frac{C_2 b_0}{C_1} \right ] , 
\nonumber \\ 
B_g^{(1)} & = & - \frac{\beta_0}{2} -
     6 \ln \frac{C_2 b_0}{C_1} ,
\nonumber
\end{eqnarray}
where $C_F = (N_c^2 - 1)/(2 N_c)$, $C_A = N_c$
($N_c = 3$ in QCD), $b_0 = 2 \exp (- \gamma_E)$ where 
$\gamma_E$ is the Euler constant, $n_f$ is the number of active quarks
(with masses $m_q < C_2 Q$), the coefficient $K$~\cite{SF-quark} is:
\begin{displaymath}
K = C_A \left [ \frac{67}{18} - \frac{\pi^2}{6} \right ] -
   \frac{5}{9} \, n_f,
\end{displaymath}
and $\beta_0$ has been specified earlier in Eq.~(\ref{eq:alpha-s}). 
Note that the coefficient $B_q^{(1)}$ given above in
Eq.~(\ref{eq:coeff}) is taken from Ref.~\cite{Stefanis99} and it 
differs from the corresponding coefficient given in Ref.~\cite{SF-gluon}.

The knowledge of the LL and NLL perturbative coefficients $A_p^{(1)}$,
$A_p^{(2)}$, and $B_p^{(1)}$ as well as the strong coupling
$\alpha_s (q^2)$ at the two-loop level~(\ref{eq:alpha-s}) allows to get
the explicit expression for the LL and NLL terms of the Sudakov functions
in the form:
\begin{eqnarray}
\hspace{-3mm}
&& 
s_p (Q, b, C_1, C_2) = \frac{A_p^{(1)}}{\beta_0} \, L_Q
\left [ \ln \lambda + \frac{1}{\lambda} - 1 \right ] 
\label{eq:S-NLL} \\ 
\hspace{-3mm} 
&& -  
\frac{4 A_p^{(2)}}{\beta_0^2}
\left [ \ln \lambda - \lambda + 1 \right ] +
\frac{ B_p^{(1)}}{\beta_0} \, \ln \lambda
+ \frac{2 \beta_1 A_p^{(1)}}{\beta_0^3}
\nonumber \\ 
\hspace{-3mm} 
&& \times 
\left [ \lambda \, \ln \lambda + (1 - \lambda) \, (1 + \ln L_Q) +
\ln L_Q \, \ln \lambda - \frac{1}{2} \ln^2 \lambda \right ] ,
\nonumber
\end{eqnarray}
where $\lambda = L_Q / L_b$, $L_Q = \ln (C_2^2 Q^2 / \Lambda^2)$, and   
$L_b = \ln (C_1^2 / b^2 \Lambda^2)$.

Let us discuss the range of the impact parameter~$b$. In general, it 
is defined as a positive variable but in the region $b \ge 1 / \Lambda$
the Sudakov exponent diverges due to the Landau pole in the QCD
coupling $\alpha_s (\mu^2)$ at $\mu = \Lambda$. In this case the
perturbative calculation is no longer valid and a prescription
for parametrizing the non-perturbative physics in the low
transverse momentum (or large $b$) region is
necessary. We shall restrict ourselves to the region $b < 1/ \Lambda$.
There also exists a lower bound on the Sudakov function, arising from
the consideration that the transverse  momentum of the emitted gluons is
not allowed to be large, i.e., we have to restrict to the region 
$|{\bf k}_\perp| \ll C_2 Q$, where $C_2 Q$ is the hard scale. Thus, the
region of applicability of the Sudakov formalism is limited to the region:
$1/C_2 Q \ll b \ll 1/ \Lambda$.

In the following it is necessary to fix the arbitrary parameters
$C_1$ and $C_2$. The constant $C_1$ determines the onset of the
non-perturbative physics. The renormalization constant $C_2$ specifies
the scale of the hard scattering process. Following
Ref.~\cite{Sudakov-gen}, we make 
the choice $C_1 = 1$ and $C_2 = 1/\sqrt 2$ in our
analysis\footnote{Another choice of this parameters is 
$C_1 = b_0$ and $C_2 = 1$~\cite{CS} which would eliminate
large constants in the functions $A_p$ and $B_p$ of
Eq.~(\ref{eq:AB-expansion}).}.

Using the definitions of the $\bar q q$ and gluonic parts 
of the $\eta^\prime$-meson vertex~(\ref{eq:FFQ-def}) 
and~(\ref{eq:FFG-def}), respectively, the results for the 
corresponding vertices, including the dependence on the 
transverse momenta, are:
\begin{eqnarray}
\hspace{-3mm}
&& 
F^{(q)}_{\egg} (q_1^2, q_2^2, m_{\eta'}^2) = 
4 \pi \alpha_s (Q^2) \, \frac{2}{\sqrt{2 N_c}} 
\label{eq:FFq-mHSA-1} \\
\hspace{-3mm} 
&& \times 
\int\limits_0^1 \! dx \!\! \int \! \frac{d {\bf b}}{4 \pi}
\hat \Psi^{(q)} (x, Q, {\bf b}) \,
S^{(q)} (x, Q, b)
\int \! \frac{d {\bf k}_\perp}{(2 \pi)^2} \,
{\rm e}^{- i ({\bf k}_\perp {\bf b})}
\nonumber \\ 
\hspace{-3mm} 
&& \times 
\left [
\frac{1}
     {x q_1^2 + \bar x q_2^2 - x \bar x m_{\eta'}^2 - {\bf k}_\perp^2
      + i \epsilon}
+ (x \leftrightarrow \bar x)
\right ],
\nonumber \\
\hspace{-3mm} 
&& 
F^{(g)}_{\egg} (q_1^2, q_2^2, m_{\eta'}^2) = 
\frac{4 \pi \alpha_s (Q^2)}{Q^2} \, \sqrt{2 N_c} 
\label{eq:FFg-mHSA-1} \\ 
\hspace{-3mm} 
&& \times 
\int\limits_0^1 \! dx \!\! \int \! \frac{d {\bf b}}{4 \pi}
\hat \Psi^{(g)} (x, Q, {\bf b}) \, S^{(g)} (x, Q, b)
\int \! \frac{d {\bf k}_\perp}{(2 \pi)^2} \,
{\rm e}^{- i ({\bf k}_\perp {\bf b})}
\nonumber \\ 
\hspace{-3mm} 
&& \times 
\left [   
\frac{m_{\eta'}^2 + (x - \bar x) (q_1^2 - q_2^2)}
     {x q_1^2 + \bar x q_2^2 - x \bar x m_{\eta'}^2 - {\bf k}_\perp^2
     + i \epsilon}
- (x \leftrightarrow \bar x) \right ] .
\nonumber
\end{eqnarray}

For the space-like gluon virtualities [$q_1^2 < 0$, $q_2^2 < 0$, and 
$Q^2 = - (q_1^2 + q_2^2)$] the integration over the transverse momentum
${\bf k}_\perp$ can be done resulting in the modified Bessel function of
order zero, $K_0 (z)$. Taking this into account and substituting the
$\eta^\prime$-meson wave-function~(\ref{eq:WF-full}), the quark and
gluonic contributions reduce to the two-dimensional integrals of the form:
\begin{widetext} 
\begin{eqnarray}
F^{(q)}_{\egg} (Q, \omega, \eta) & = &
- 4 \pi \alpha_s (Q^2) \, \frac{C}{N_c} \int\limits_0^1 dx \,
\phi^{(q)} (x, Q)
\int\limits_0^\infty db \, b \, {\rm e}^{- x \bar x b^2 / 4 a^2} \,
S^{(q)} (x, Q, b) \, K^{(+)}_0 (x, b Q)
\label{eq:FFq-mHSA-2} \\ 
& = &
- 4 \pi \alpha_s (Q^2) \, \frac{C}{N_c \Lambda^2} \int\limits_0^1
dx \, \phi^{(q)} (x, Q)
\int\limits_0^1 db_\Lambda \, b_\Lambda \,
\exp \left [ - \frac{x \bar x}{4 a^2 \Lambda^2} \, b_\Lambda^2 \right ] \,
S^{(q)} ( x, Q, b) \, K_0^{(+)} (x, b_\Lambda Q_\Lambda) ,
\nonumber \\
F^{(g)}_{\egg} (Q, \omega, \eta) & = &
- 4 \pi \alpha_s (Q^2) \, C 
\int\limits_0^1 dx \, \phi^{(g)} (x, Q)
\int\limits_0^\infty db \, b \,
{\rm e}^{- x \bar x b^2 / 4 a^2} \, S^{(g)} (x, Q, b) 
\label{eq:FFg-mHSA-2} \\
& \times &
\left [ |\eta| \, K^{(-)}_0 (x, b Q) -
(x - \bar x) \, \omega \, K^{(+)}_0 (x, b Q) \right ]
\nonumber \\
& = &
- 4 \pi \alpha_s (Q^2) \, \frac{C}{\Lambda^2}
\int\limits_0^1 dx \, \phi^{(g)} (x, Q)
\int\limits_0^1 db_\Lambda \, b_\Lambda \,
\exp \left [ - \frac{x \bar x}{4 a^2 \Lambda^2} \, b_\Lambda^2 \right ] \,
\nonumber \\
& \times &
S^{(g)} (x, Q, b) \,
\left [ |\eta| K_0^{(-)} (x, b_\Lambda Q_\Lambda) -
(x - \bar x) \omega K_0^{(+)} (x, b_\Lambda Q_\Lambda) \right ] ,
\nonumber 
\end{eqnarray}
where
\[
K^{(\pm)}_0 (x, b Q) = K^{(\pm)}_0 (x, b_\Lambda Q_\Lambda) =
\frac{1}{2} \left [
K_0 \left ( b_\Lambda Q_\Lambda \lambda_+ (x, \omega, \eta) \right )
\pm K_0 \left ( b_\Lambda Q_\Lambda \lambda_+ (\bar x, \omega, \eta)
\right )
\right ] ,  
\]
\end{widetext}
with the dimensionless parameters defined as $b_\Lambda = b \Lambda$,
$Q_\Lambda = Q / \Lambda$, $|\eta| = m_{\eta^\prime}^2 /Q^2$; 
the function $\lambda_\pm^2 (x, \omega, \eta) =
[1 + \omega (x - \bar x) \pm 2 x \bar x |\eta|]/2$,
and the quark and gluonic components of the $\eta^\prime$-meson 
wave-function are taken in the form of Eq.~(\ref{eq:WF-full}).
The restriction $b \le 1/\Lambda$ due to the Sudakov formalism is 
also taken into account in the final result for the space-like gluons 
presented in Eqs.~(\ref{eq:FFq-mHSA-2}) and~(\ref{eq:FFg-mHSA-2}). 

\begin{figure}[tb]
%
\psfig{file=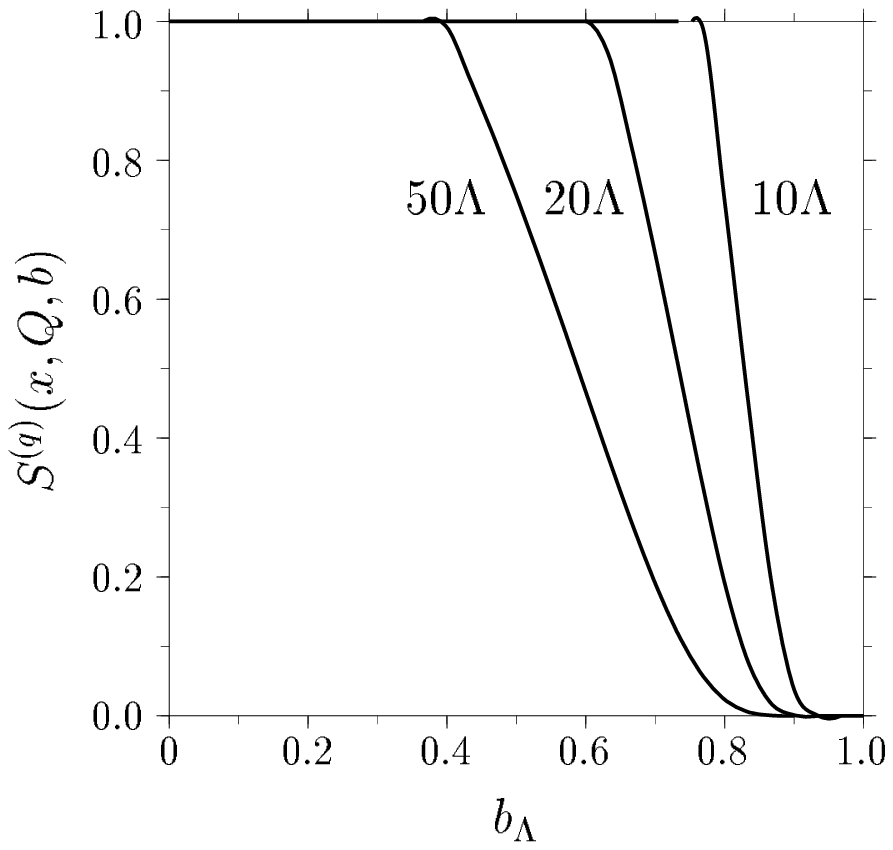,viewport=120 410 420 670,width=.45\textwidth} 
\psfig{file=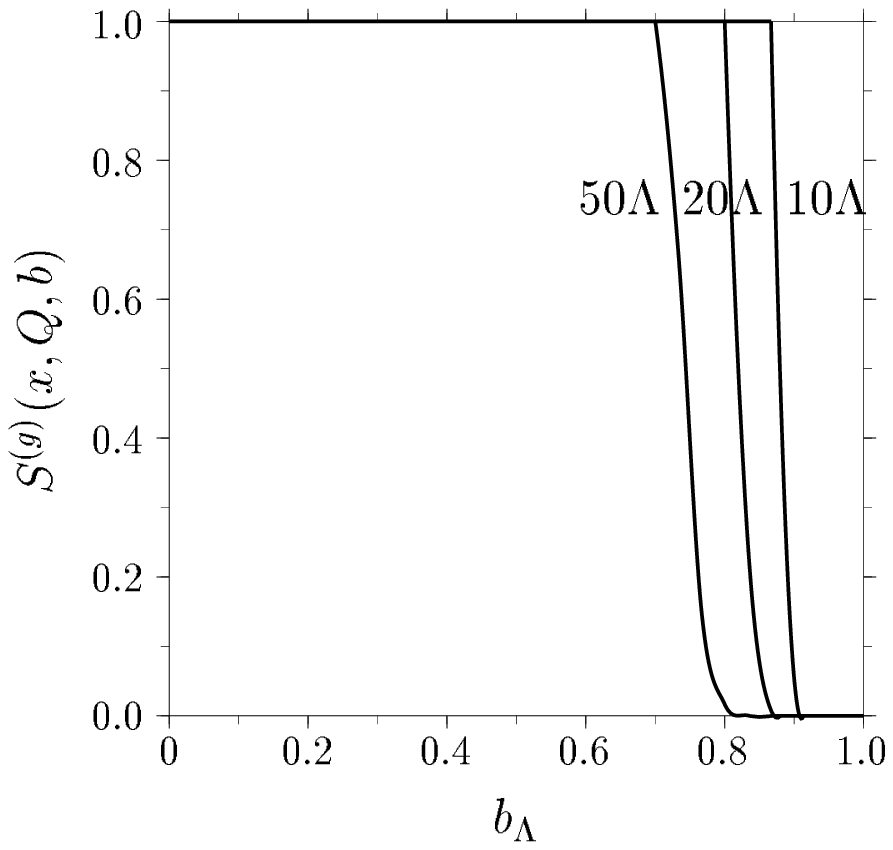,viewport=120 410 420 670,width=.45\textwidth}
\caption{\label{fig:sud-fac}%
         The Sudakov function plotted in terms of the dimensionless 
         impact parameter~$b_\Lambda$ for quarks (upper figure) 
         and gluons (lower figure) with $x = 1/2$
         and $Q = 10\Lambda$, $20\Lambda$, and~$50\Lambda$.}
\end{figure}

The arguments of the Bessel functions 
$K_0 (b Q \lambda_+ (x, \omega, \eta))$ 
and $K_0 (b Q \lambda_+ (\bar x, \omega, \eta))$
in the $\eta^\prime$-meson vertex 
define the largest scale parameter~$t$ in the Sudakov
factor~(\ref{eq:S-meson}):
\begin{equation}
t = \max \left \{
Q \lambda_+ (x, \omega, \eta),
Q \lambda_+ (\bar x, \omega, \eta), 1/ b
\right \}.
\label{eq:max-scale}
\end{equation}
The dependence of the Sudakov factors~$S^{(q)} (x, Q, b)$
and~$S^{(g)} (x, Q, b)$ on the impact parameter~$b$ at $x = 1/2$ 
and $Q = 10 \Lambda$, $20 \Lambda$ and
$50 \Lambda$ is presented in Fig.~\ref{fig:sud-fac}.
%
%
%
%
It is seen that the Sudakov factors give a cut-off in addition 
to the restriction $b \le 1 / \Lambda$ of the cut-off of the 
integration interval over~$b$. With increasing the typical mass 
scale~$Q$ this effect is stronger in the quark contribution. 

\begin{figure}[tb]
%
\psfig{file=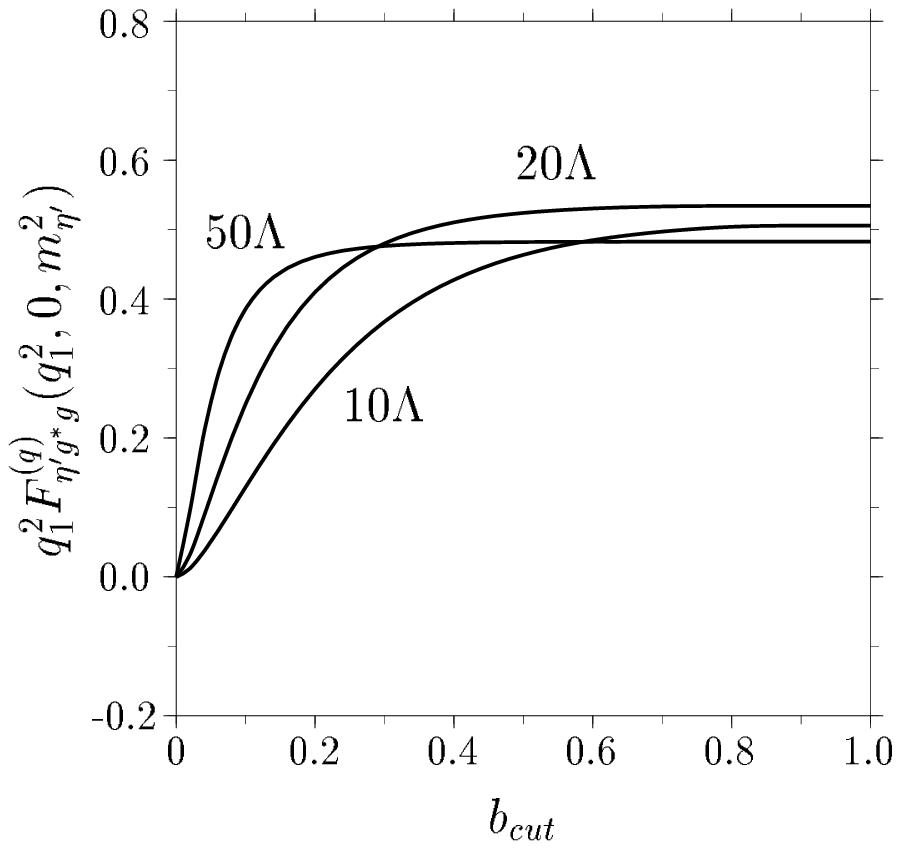,viewport=120 370 420 630,width=.45\textwidth} 
\psfig{file=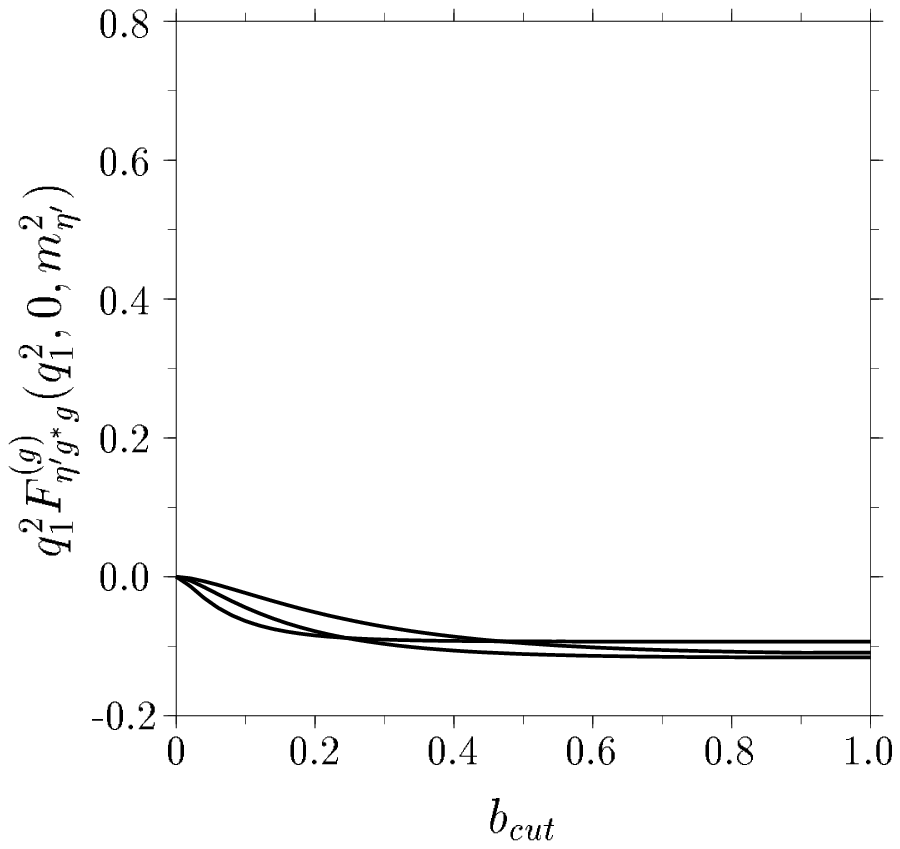,viewport=120 370 420 630,width=.45\textwidth}
\caption{\label{fig:ff-sud-bm}%
         The quark-antiquark (upper figure) and gluonic (lower 
         figure) contributions to the $\eta^\prime g^* g$ vertex function
         $q_1^2 F^{(p)}_{\eta^\prime g^* g}   
         (q_1^2, 0, m_{\eta^\prime}^2)$ versus the cut off in the impact
         parameter space,~$b_{cut}$,  
         for the space-like gluon virtuality $q_1^2 = - Q^2$ 
         with $Q = 10 \Lambda$, $20 \Lambda$, and~$50 \Lambda$.} 
\end{figure}

In Fig.~\ref{fig:ff-sud-bm} we show the distribution of the quark
and gluonic parts of the vertex in the impact parameter
space integrating here with a variable cut off, $b_{cut}$. The curves,
showing the dependence of $q_1^2 \, F^{(p)}_{\eta^\prime g^* g} 
(q_1^2, 0, m_{\eta^\prime}^2)$ on $b_{cut}$, start from zero for 
$b_{cut} = 0$ and reach their full height for $b_{cut} = 1$ beyond which
we consider any remaining contributions as non-perturbative.
%
%
%
%
The faster rise of the $\egg$ vertex is observed for the quark
contribution as~$Q$ increases while the gluonic one is mildly sensitive
to the typical mass scale. At the largest value shown, $Q = 50 \Lambda$,
the quark curve is quite flat for $b_{cut} \ge 0.2$ which
indicates a small contribution from this region. As for the smaller
values, for example, $Q = 10 \Lambda$, the vertex gets
a rather large contribution from large distances in the impact parameter
space. The gluonic part of the vertex receives the main
contribution from the short distances $b \le 0.3$. In contrast to 
the quark contribution, the gluonic contribution is negative and 
decreases the summed vertex function by approximately 20\%.    

To obtain the $\eta^\prime$-meson vertex function for 
the time-like gluon virtualities one has to replace $Q^2 \to - Q^2$ 
and $|\eta| \to - \eta$ in Eqs.~(\ref{eq:FFq-mHSA-2}) 
and~(\ref{eq:FFg-mHSA-2}). The vertex function is an analytic 
function of the scale parameter~$Q^2$ and it can be 
analytically continued from the space-like~$Q^2$ region to the 
time-like one. As mentioned earlier,
the $\egg$ vertex results from convoluting the $\eta^\prime$-meson
wave-function and the hard scattering amplitude kernel. As we do not
take the QCD corrections to the hard scattering amplitude into account,
the strong coupling, $\alpha_s (q^2)$, has the same dependence on the 
scale parameter in both the space-like and time-like regions  
($\alpha_s (- q^2) = \alpha_s (q^2) [1 + {\cal O} (\alpha_s)]$). 
The other component of the $\eta^\prime$-meson wave-function 
sensitive to the transition from the space-like region to the
time-like one is the Sudakov factor~(\ref{eq:S-meson}). 
The analytic continuation of the Sudakov factor was analyzed in
Refs.~\cite{MS90,Gousset} with the result that the time-like and 
space-like form factors have the same scale dependence 
in the asymptotic regime. We use the space-like expression for the
Sudakov factor to study the $\egg$ vertex for the
time-like gluon virtualities. The hard scattering amplitude kernel 
gives the additional transfer momentum dependence due to the quark
and gluon propagators. We recall that, in general, the propagator 
is defined in the complex momentum plane and, hence, it can have
both real and imaginary parts in the time-like region of the momentum
due to $i \epsilon$ term in the denominator. Here, we perform the
analysis of the propagators in the transverse momentum space keeping
their transverse momenta dependence. In mHSA it is more convenient to
operate with the Fourier transform of the propagators in the impact
parameter space but this analysis can be easily reformulated
in terms of the impact parameter~${\bf b}$, which is the conjugate
variable to~${\bf k}_\perp$.  

\begin{figure}[tb]
%
\psfig{file=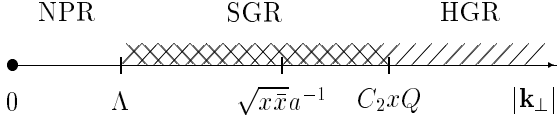,viewport=150 600 470 680,width=.45\textwidth}
\caption{\label{fig:k-perp}%
         Different regions in the integration
         plane in the variable~${\bf k}_\perp$.
         The abbreviations used correspond to non-perturbative (NPR),
         soft-gluon (SGR), and hard-gluon (HGR) regions.
         The point $|{\bf k}_\perp| = 0$ corresponds to
         the Brodsky-Lepage approach.}
\end{figure}

The ${\bf k}_\perp$ integration plane can be divided in several 
regions as shown in Fig.~\ref{fig:k-perp}. The interval
$0 < |{\bf k}_\perp| < \Lambda$ is the non-perturbative region.
We note that the Brodsky-Lepage approach, in which  
the dependence on the transverse momentum is neglected, corresponds to the
 point ${\bf k}_\perp = 0$ in Fig.~\ref{fig:k-perp}. It is seen that
this transverse momentum value lies in the deeply non-perturbative region
and too far from the contributing region of the transverse momentum
integration. It means that in the Brodsky-Lepage approach the imaginary
part of the $\egg$ vertex can not be calculated correctly
and the natural prescription is to drop it altogether. Another consequence
of the approximate forms of the propagators in the Brodsky-Lepage
picture is the appearance of the singularity in the $\egg$ vertex
in the region close to the $\eta^\prime$-meson mass.
The interval $|{\bf k}_\perp| > \Lambda$ can be
effectively divided into two regions: the soft-gluon region (SGR)
where the Sudakov factor is taken into account and the hard-gluon region
(HGR) where the Sudakov effect is absent. In Fig.~\ref{fig:k-perp} we
also point out the value of $\sqrt{x \bar x} a^{-1}$ to show the
importance of the transverse momentum distribution of the
$\eta^\prime$-meson wave-function. In the region
$|{\bf k}_\perp| > \sqrt{x \bar x}a^{-1}$ the contribution to the
vertex is effectively suppressed by the assumed Gaussian behavior
of the transverse momentum distribution~(\ref{eq:WF-full}).
%
%
%
%

To calculate the integrals~(\ref{eq:FFq-mHSA-1}) and~(\ref{eq:FFg-mHSA-1}) 
over the transverse momentum~${\bf k}_\perp$ in the time-like region of
the gluon virtualities it is useful to do the analytic continuation
of~$|{\bf k}_\perp|$ in the complex plane. The new feature with respect
to the space-like case is that the contour of the transverse momentum
integration goes now near the pole located at
${\bf k}_\perp^2 = Q^2 \, \lambda_-^2 + i \epsilon$. 
This pole, except in the end point regions [$x \to 0$ for the second 
gluon on the mass shell, or $x \to 1$ for the first one due to the 
function $\lambda_- (x, \omega, \eta)$], is far from the bounds of
integration of the variable $|{\bf k}_\perp|$. Therefore, this integral
can be evaluated by deforming the contour of the integration in the
complex plane of this variable.

The pole encountered in our computations when an internal quark or gluon 
goes on the mass shell do not correspond to an observable state. 
They appear in the effective $\egg$ vertex which is not an observable 
by itself. Provided this vertex is included in the complete 
amplitude of some real process such as $B \to \eta^\prime K$ and 
$B \to \eta^\prime X_s$, this amplitude will have an additional complex
phase factor as a reminder of this singularity. Notice that the complete
physical amplitude of the process including the effective $\egg$ vertex
in a purely hadronic computations has also the poles reflecting the
existence of the intermediate physical (mass-shell) states which are
hadronic ones. Therefore, it is natural to expect the appearance of the
additional phase factor here also.  
 
\begin{figure}[tb]
%
\psfig{file=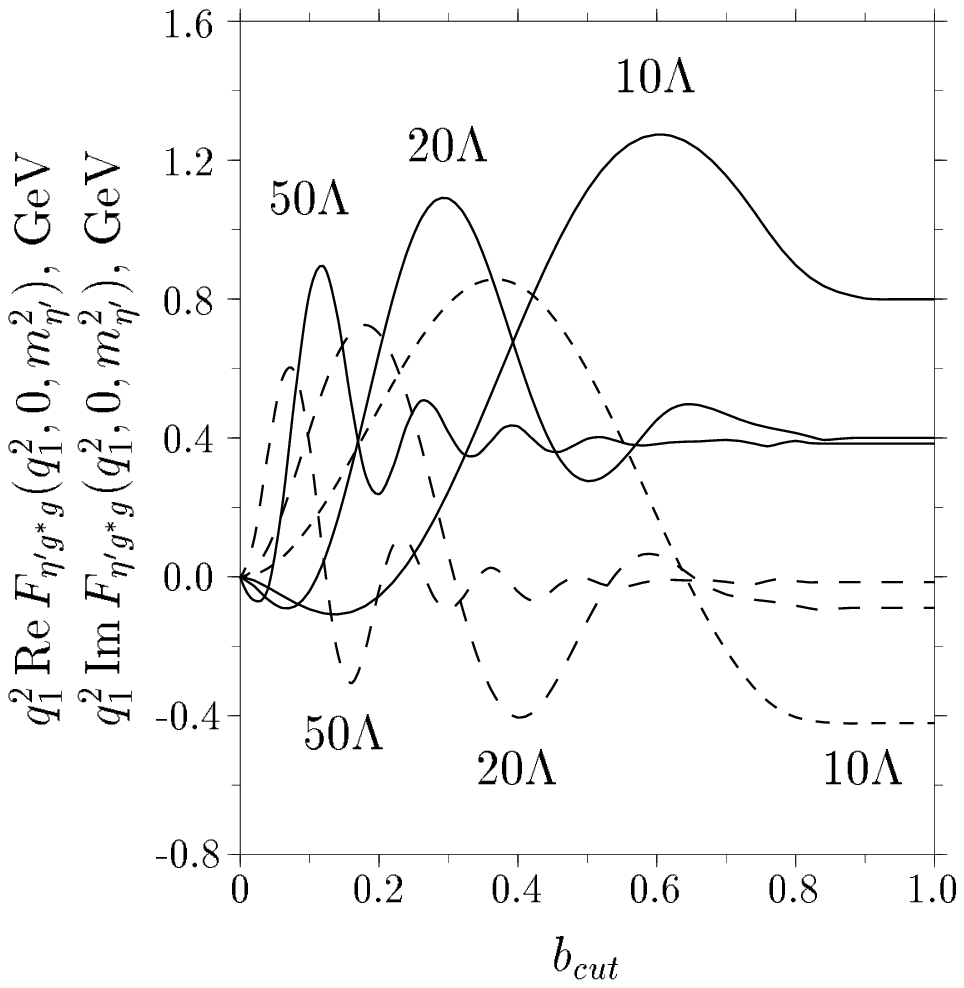,viewport=120 320 410 620,width=.45\textwidth} 
\psfig{file=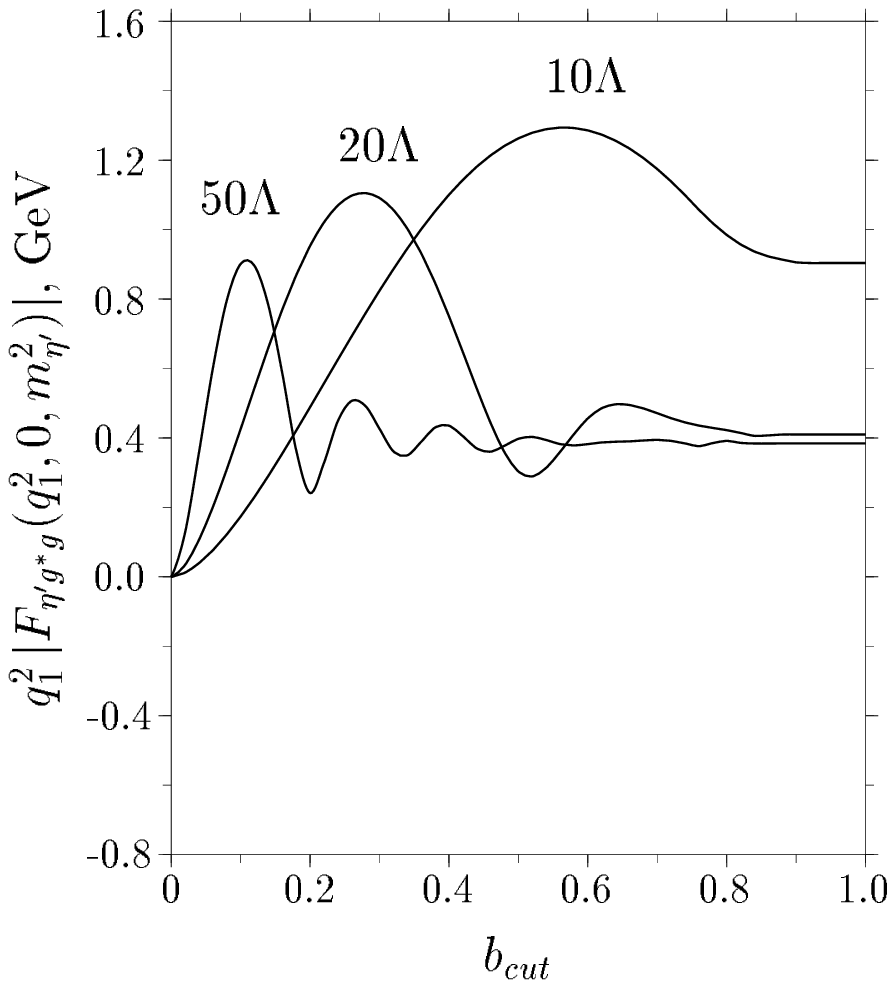,viewport=120 320 410 620,width=.45\textwidth}
\caption{\label{fig:ff-sud-bp}%
         The real (solid curves in the upper figure) and imaginary parts
         (dashed curves in the upper figure) and the absolute
         value of the vertex function $q_1^2 F_{\eta^\prime g^* g}
         (q_1^2, 0, m_{\eta^\prime}^2)$ (lower figure), plotted as
         functions of the cut off,~$b_{cut}$, in the impact parameter
         space for the time-like gluon virtuality, $q_1^2 = Q^2$, 
         with $Q = 10\Lambda$, $20\Lambda$, and~$50\Lambda$.}
\end{figure}

According to the arguments given above, the $\egg$ vertex 
for the time-like gluon virtualities can be obtained from
Eqs.~(\ref{eq:FFq-mHSA-2}) and~(\ref{eq:FFg-mHSA-2}) by changing 
$|\eta| \to - \eta$ and $K_0 (b Q \lambda_+ (x, \omega, \eta)) \to 
K_0 (i b Q \lambda_- (x, \omega, \eta))$. If one uses the relation
$K_0 (i z) = - i \pi \, H_0^{(2)} (z)/ 2$ where
$H_0^{(2)}(z) = J_0 (z) - i Y_0 (z)$,
(here, $H_0^{(2)} (z)$, $J_0 (z)$ and $Y_0 (z)$ are the
second Hankel function, Bessel function of the first kind, and Neumann
function, respectively~\cite{Bateman}), it is seen that in contrast to 
the case of the space-like gluon virtualities the vertex
gets an imaginary part. After these changes the quark and gluonic 
contributions to the $\egg$ vertex function are:
\begin{widetext}
\begin{eqnarray}
F^{(q)}_{\egg} (Q, \omega, \eta) & = &
4 \pi \alpha_s (Q^2) \, \frac{i \pi C}{2 N_c \Lambda^2}
\int\limits_0^1 dx \, \phi^{(q)} (x, Q)
\int\limits_0^1 db_\Lambda \, b_\Lambda \,
\exp \left [ - \frac{x \bar x}{4 a^2 \Lambda^2} \, b_\Lambda^2 \right ] \,
S^{(q)} ( x, Q, b) \, H_0^{(+)} (x, b_\Lambda Q_\Lambda) ,
\qquad 
\label{eq:FFq-mHSA-3} \\
F^{(g)}_{\egg} (Q, \omega, \eta) & = &
4 \pi \alpha_s (Q^2) \, \frac{i \pi C}{2 \Lambda^2}
\int\limits_0^1  dx \, \phi^{(g)} (x, Q)
\int\limits_0^1 db_\Lambda \, b_\Lambda \,
\exp \left [ 
- \frac{x \bar x}{4 a^2 \Lambda^2} \, b_\Lambda^2 \right ] 
\qquad
\label{eq:FFg-mHSA-3} \\
& \times &
S^{(g)} ( x, Q, b ) \,
\left [
\eta H_0^{(-)} (x, b_\Lambda Q_\Lambda)
+ (x - \bar x) \omega H_0^{(+)} (x, b_\Lambda Q_\Lambda)
\right ] ,
\nonumber 
\end{eqnarray}
where
\[
H_0^{(\pm)} (x, b Q) = H_0^{(\pm)} (x, b_\Lambda Q_\Lambda) =
\frac{1}{2} \left [
H^{(2)}_0 
\left (b_\Lambda Q_\Lambda \lambda_- (x, \omega, \eta ) \right )
\pm
H^{(2)}_0 
\left ( b_\Lambda Q_\Lambda \lambda_- (\bar x, \omega, \eta) \right )
\right ] .
\]
\end{widetext}
The real and imaginary parts of the $\egg$ vertex in the form
$q_1^2 \, F^{(p)}_{\eta^\prime g^* g}$
are shown as functions of the impact parameter cut off,
$b_{cut}$, in Fig.~\ref{fig:ff-sud-bp}.
%
%
%
%
As in the case of the space-like virtualities, a fast rise is observed 
with~$Q$ in the region of small values of the
impact parameter cut-off. The oscillatory behavior of the cylindric 
functions results in an oscillatory form of the vertex function for small 
and moderate values of the impact parameter. The contribution to the
vertex from the large distances, $b \ge 0.8$, is small. It is also
seen that as the scale parameter~$Q$ increases, the relative 
contribution of the imaginary part in the absolute value of the vertex
is strongly suppressed, and the vertex is mainly defined 
by the real contribution in the large~$Q^2$ asymptotics.  

The asymptotics of the quark part of the vertex
$F_{\egg}^{(q)}$ in the limit of large~$Q^2$ is:
\begin{eqnarray}
F_{\egg}^{(q)} & \simeq & 4 \pi \alpha (Q^2) \, \frac{3 C}{N_c Q^2} \, 
\left \{ f_0 (\omega) + 
\frac{2 i \pi}{\omega} \, y_+ y_- 
\right. 
\label{eq:FFq-mHSA-asymp} \\
& \times & 
\left. 
\vphantom{\frac{2 i \pi}{\omega} \, y_+ y_- \,} 
(1 - y_+ y_-)   
\left [
H_0^{(2)} (Q_\Lambda y_+) - H_0^{(2)} (Q_\Lambda y_-) 
\right ] \right \}, 
\nonumber 
\end{eqnarray}  
where $y_{\pm} = \sqrt{(1 \pm \omega)/2}$,
and the function $f_0 (\omega)$ is defined in Eq.~(\ref{eq:f-func-asymp}).
In this expression the large argument asymptotics of the Hankel function 
should be used: $H_0^{(2)} (z) \simeq \sqrt{2 / \pi z} \, 
\exp (- i [z - \pi / 4])$~\cite{Bateman}.   
The same correction of order $1 / Q^{5/2}$ to the gluonic part of the
vertex function $F_{\egg}^{(g)}$ is equal to zero and one can use
the asymptotic behavior defined by Eq.~(\ref{eq:GFF-asymp}).

\section{\label{sec:numeric}%
         Numerical Analysis and Comparison with Existing Results}

In this section we give a numerical analysis of the $\egg$ vertex 
in the time-like and space-like regions, derived in
the preceding sections. To that end we specify the input parameters.
The dimensional factor~$C$ [see Eq.~(\ref{eq:C-const})] contains the 
decay constants~$f_q$ and~$f_s$ and the mixing angle~$\phi$ which 
can be constrained from the existing experimental data.
We shall adopt here the Feldmann-Kroll-Stech $\eta - \eta^\prime$ mixing
scheme~\cite{FKS98}, which is 
phenomenologically consistent and satisfies the constraints from chiral
perturbation theory \cite{Leutwyler}, yielding~\cite{FKS98}: 
$f_q = (1.07 \pm 0.02) \, f_\pi$, $f_s = (1.34 \pm 0.06) \, f_\pi$, 
and $\phi = 39.3^\circ \pm 1.0^\circ$, where $f_\pi \simeq 131$~MeV 
is the pion decay constant. Using the central values of the parameters, 
one gets $C \simeq 2 f_\pi \simeq 260$~MeV. 

The quark and gluonic wave-functions~(\ref{eq:qef-gen}) 
and~(\ref{eq:gef-gen}) contain both free ($B^{(q)}_n$, $B^{(g)}_n$) 
and constrained ($\rho^{(q)}_n$, $\rho^{(g)}_n$) parameters. 
The constrained parameters depend on the anomalous dimensions and are
given in Appendix~\ref{app:evol-eqns},  
while the free parameters can be fitted from the experimental data, 
for example, from the $\eta^\prime \gamma^* \gamma$ transition form
factor. We take the following restrictions on the first correction to the 
leading order wave-function~(\ref{eq:qef-gen}): $|B^{(q)}_2| < 0.1$ and  
$|\rho^{(g)}_2 \, B^{(g)}_2| < 0.1$ in order to keep $\phi^{(q)} (x, Q)$ 
close to its asymptotic value: $\phi_{\rm as} (x) = 6 x \bar x$, in 
agreement with the experimental data on the $\eta^\prime \gamma^* \gamma$
transition form factor~\cite{eta'-gamma}. Taking into account 
$\rho^{(g)}_2 \simeq - 1/90$ from Eq.~(\ref{eq:num-values})  
we get\footnote{This differs considerably
from the values used in~\cite{Muta}.} $|B^{(g)}_2| < 9.0$.
Below, we shall present the $\eta^\prime g^* g^*$ vertex function
for the maximum allowed values of the non-perturbative parameters,
i.e., $|B^{(q)}_2| = 0.1$ and $|B^{(g)}_2| = 9.0$. We have also studied 
numerically the resulting vertex function for 
considerably smaller values of these parameters. In that context we note
that as the NLO quark contribution to the overall $\eta^\prime g^* g^*$
vertex function is small, there is not much sensitivity to the variation 
of the parameter~$B^{(q)}_2$. Hence, we fix this parameter to its maximum
allowed value $|B^{(q)}_2| = 0.1$. However, there is considerable
sensitivity to the variation of the parameter~$B^{(g)}_2$. To show this
we shall take $|B^{(g)}_2| = 3.0$. The resulting
theoretical dispersion between the two cases ($|B^{(g)}_2| = 9.0$ vs.
$|B^{(g)}_2| = 3.0$) can be taken as representative of the theoretical 
uncertainties due to the undetermined non-perturbative parameters.   

\begin{figure}[tb]
%
\psfig{file=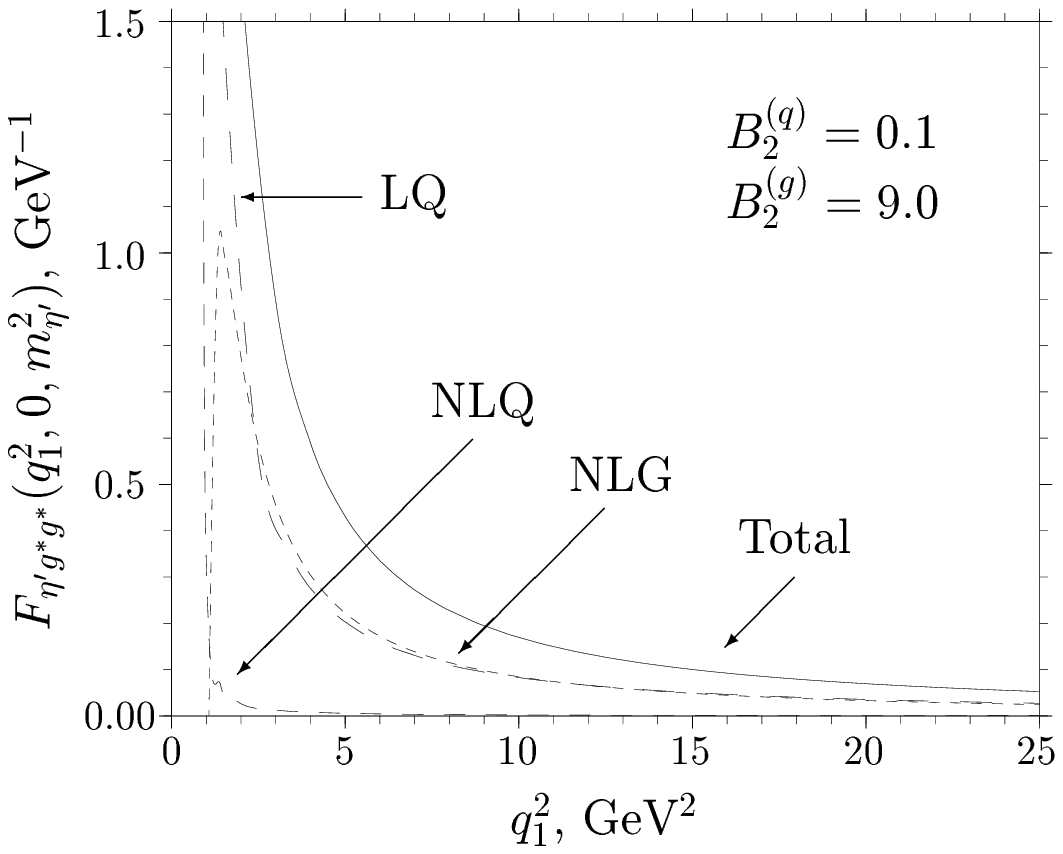,viewport=120 445 460 710,width=.45\textwidth} 
\psfig{file=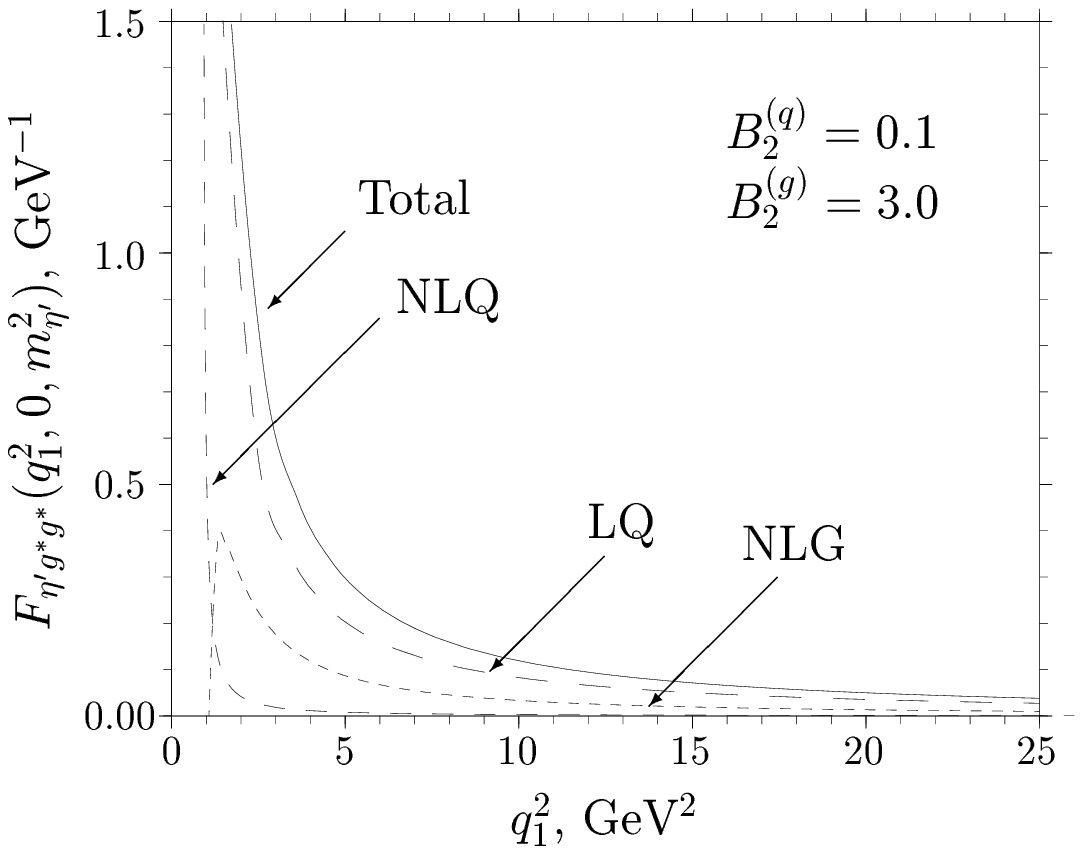,viewport=120 445 460 710,width=.45\textwidth} 
\caption{\label{fig:form-fac}%
         The $\eta^\prime g^* g$ vertex
         $F_{\egg} (q_1^2,0,m_{\eta^\prime}^2)$ as a function 
         of~$q_1^2$ with $B^{(q)}_2 = 0.1$ and two values of~$B^{(g)}_2$:  
         $B^{(g)}_2 = 9.0$ (the upper plot) and $B^{(g)}_2 = 3.0$ (the 
         lower plot) in the Brodsky-Lepage approach.
         The dashed curves are the leading (LQ), next-to-leading 
         quark-antiquark (NLQ), and gluonic (NLG) components, and the
         solid curve is the sum.}  
\end{figure}

First, we consider the case when one of the gluons 
is on the mass shell, which we take for the sake of
definiteness to be the second one ($q_2^2 = 0$). 
In Fig.~\ref{fig:form-fac} we show the leading (long-dashed curve) 
and next-to-leading (middle-dashed curve) quark contributions 
as well as the gluonic one (short-dashed curve) for the time-like 
gluon virtuality ($q_1^2 > m_{\eta^\prime}^2$) corresponding to the 
maximal values of the free parameters: $B^{(q)}_2 = 0.1$ and 
$B^{(g)}_2 = 9.0$ (upper plot), and for $B^{(q)}_2 = 0.1$
and $B^{(g)}_2 = 3.0$ (lower plot). The
solid curve in each of these figures is the total contribution to
$F_{\eta^\prime g^* g^*} (q_1^2, 0, m_{\eta^\prime}^2)$ in next-to-leading
order. 
%
%
%
%
One can see from this figure that with the given parameters, 
the next-to-leading quark contribution is subdominant 
in the entire region of~$Q^2$ shown, except in the neighborhood 
of the threshold $Q^2 = m_{\eta^\prime}^2$, where both the leading
and non-leading quark contributions have logarithmic divergences. 
Notice that in this region our results are not valid and we can trust 
only the region $Q^2 \gg m_{\eta^\prime}^2$. The gluonic contribution 
(called NLG) for the maximum values of the free parameters is
comparable to the leading quark contribution (called LQ), as shown in
the upper plot in Fig.~\ref{fig:form-fac}, increasing the
$\eta^\prime g^* g$ vertex by almost a factor 2 as compared to the 
case when only the quark content of the $\eta^\prime$-meson is assumed. 
This may be considered as the maximum gluonic content
of the $\eta^\prime$-meson allowed by current data. Even in the
more realistic case with $B^{(q)}_2 = 0.1$ and $B^{(g)}_2 = 3.0$, we
see from the lower plot in Fig.~\ref{fig:form-fac} that the
gluonic contribution  is not small, and also in this case it enhances
the value of the total $\eta^\prime g^* g$ vertex function (the solid 
curve in Fig.~\ref{fig:form-fac}) at the level of few tens percent. 

\begin{figure}[tb]
%
\psfig{file=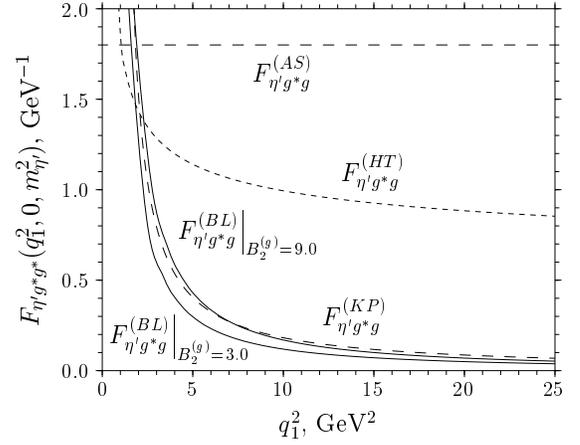,viewport=125 445 460 710,width=.45\textwidth} 
\caption{\label{fig:contribp}%
         The $\eta^\prime g^* g$ vertex for the
         time-like gluon virtuality with $B^{(q)}_2 = 0.1$ and 
         two values of~$B^{(g)}_2$: $B^{(g)}_2 = 9.0$ (upper solid 
         curve) and $B^{(g)}_2 = 3.0$ (lower solid curve) calculated 
         in the Brodsky-Lepage approach in this paper.  
         The long- and short-dashed curves correspond to the functions 
         $F^{(KP)}_{\eta^\prime g^* g} (q_1^2, 0, m_{\eta^\prime}^2)
       = 1.8~{\rm GeV}^{-1} / (q_1^2 / m_{\eta^\prime}^2 - 1)$~\cite{TFF-1} 
         and $F^{(HT)}_{\eta^\prime g^* g} (Q^2) = 
         \sqrt 3 \alpha_s (Q^2) / (\pi f_\pi)$~\cite{TFF-2}, respectively.
         A constant form of the vertex function 
         $F^{(AS)}_{\eta^\prime g^* g} = 1.8~{\rm GeV}^{-1}$ 
         suggested in Ref.~\cite{TFF-3} is also shown as a dashed line.} 
\end{figure}

Let us compare the $\eta^\prime g^* g$ vertex calculated
here with the two set of values $B^{(q)}_2 = 0.1$ and $B^{(g)}_2 = 9.0$
(the upper solid curve in Fig.~\ref{fig:contribp}) and
$B^{(q)}_2 = 0.1$ and $B^{(g)}_2 = 3.0$ (the lower solid curve in
 Fig.~\ref{fig:contribp}) with the ones given in the literature. 
%
%
%
%
In Ref.~\cite{TFF-1}, Kagan and Petrov have parametrized the $\eta^\prime
g^* g$ vertex in the  form: 
$F_{\eta^\prime g^* g} (q_1^2, 0, m_{\eta^\prime}^2) =
H (0, 0, m_{\eta^\prime}^2) / (q_1^2 / m_{\eta^\prime}^2 - 1)$,
where $H (0, 0, m_{\eta^\prime}^2) \simeq 1.8~{\rm GeV}^{-1}$
is a phenomenological parameter extracted from the experimental data.
Its behavior is described by the long-dashed curve in
Fig.~\ref{fig:contribp}, labeled as $F_{\eta^\prime g^* g}^{(KP)}$ . 
In Ref.~\cite{TFF-2}, Hou and Tseng have parametrized the
$\eta^\prime g^* g$ vertex as: 
$F_{\eta^\prime g^* g} (q_1^2) = \sqrt 3 \alpha_s (q_1^2) /
(\pi f_\pi)$, which is drawn as the short-dashed curve 
in Fig.~\ref{fig:contribp}, labeled as $F_{\eta^\prime g^* g}^{(HT)}$.
Finally, we also show the constant vertex function 
$F_{\eta^\prime g^* g} (q_1^2) = F_{\eta^\prime g^* g}^{(AS)} = 
1.8~{\rm GeV}^{-1}$, assumed by Atwood and Soni in
Ref.~\cite{TFF-3}. It is seen that our result (the upper solid
curve) compares well
with the form  used in Ref.~\cite{TFF-1} for a wide range of~$q_1^2$, 
but is significantly different (smaller) than the one assumed 
in Ref.~\cite{TFF-2} and is in complete disagreement with the
one assumed in Ref.~\cite{TFF-3}. Our analysis shows
that in the case of the lower possible values of the  
free parameters: $B^{(q)}_2 = - 0.1$ and $B^{(g)}_2 = - 9.0$, 
the vertex function $F_{\egg}(q_1^2,0,m_{\eta^\prime}^2)$
decreases and a substantial disagreement 
with the vertex suggested in Ref.~\cite{TFF-1} would result.
However, as discussed below, data on the electromagnetic transition 
form factor of the $\eta^\prime$-meson disfavors this choice of the 
parameters.

\begin{figure}[tb]
%
\psfig{file=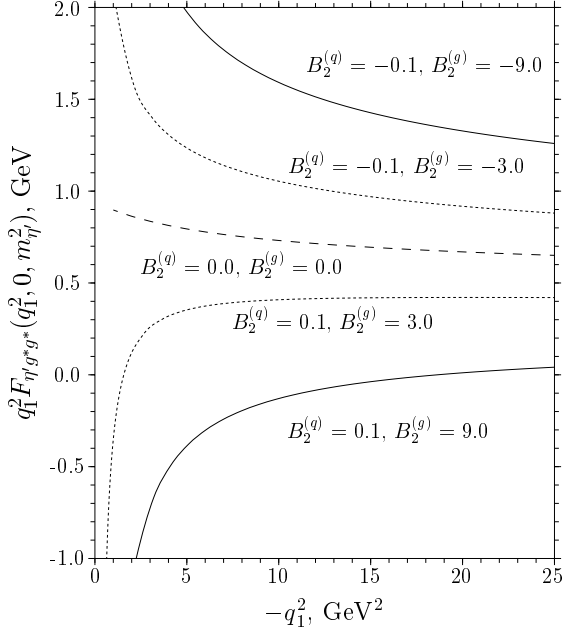,viewport=130 350 460 710,width=.45\textwidth} 
\caption{\label{fig:contribm}%
         The $\eta^\prime g^* g$ vertex function 
         $q_1^2 \, F_{\eta^\prime g^* g} (q_1^2)$ for the space-like
         gluon virtuality with $B^{(q)}_2 = - 0.1$, $B^{(g)}_2 = - 9.0$  
         and $B^{(q)}_2 = 0.1$, $B^{(g)}_2 = 9.0$ (the upper and lower 
         solid curves) and $B^{(q)}_2 = - 0.1$, $B^{(g)}_2 = - 3.0$
         and $B^{(q)}_2 = 0.1$, $B^{(g)}_2 = 3.0$ (the upper and lower 
         dotted curves) in the Brodsky-Lepage approach.   
         The dashed curve is the leading order contribution to the
         function obtained by setting $B^{(q)}_2 = B^{(g)}_2 = 0$.}
\end{figure}

We show the results of our calculations for the function 
$q_1^2 \, F_{\eta^\prime g^* g} (q_1^2)$ in the space-like region 
of the gluon virtuality ($q_1^2 = - Q^2 < 0$) in Fig.~\ref{fig:contribm}.   
%
%
%
%
The dashed curve in the middle is the leading contribution to the
$\eta^\prime g^* g$ vertex, obtained by setting the non-asymptotic 
parameters $B_2^{(q)}$ and $B_2^{(g)}$ equal to zero.
The upper and lower solid curves  are the vertex functions 
$F_{\eta^\prime g^* g} (q_1^2, 0, m_{\eta^\prime}^2)$ taking into account
the next-to-leading correction with $B^{(q)}_2 = - 0.1$,
$B^{(g)}_2 = - 9.0$ (upper curve) and $B^{(q)}_2 = 0.1$,
$B^{(g)}_2 = 9.0$ (lower curve). The dotted curves  
correspond to the choice $B^{(q)}_2 = - 0.1$, $B^{(g)}_2 = - 3.0$
(upper curve) and $B^{(q)}_2 = 0.1$, $B^{(g)}_2 = 3.0$ (lower curve).
We note  that the correction
is mainly determined by the value of the parameter $B^{(g)}_2$.
It is seen that at
small values of the gluon virtuality ($Q^2 < 5$~GeV$^2$) the complete
correction becomes approximately equal to or even larger than the leading
order contribution. This implies that in the region $Q^2 <$
few GeV$^2$  our approximation is not valid and the contributions of the
next higher order corrections become important. The behavior of
$F_{\eta^\prime g^* g} (Q^2, 0, m_{\eta^\prime}^2)$ 
for large~$Q^2$ is in  qualitative agreement with the 
corresponding electromagnetic transition form factor, 
$F_{\eta^\prime \gamma^* \gamma} (Q^2, 0, m_{\eta^\prime}^2)$,
measured in $\gamma \gamma^*$-collisions, more recently by the CLEO
and L3 collaborations~\cite{eta'-gamma}. Hence, this would suggest that 
for the free parameters  we should use the positive values: 
$B^{(q)}_2 = 0.1$ and $B^{(g)}_2 = 3.0$, or values close to them, yielding
the lower dotted curve in Fig.~\ref{fig:contribm}.
%
%
\begin{figure}[tb]
%
\psfig{file=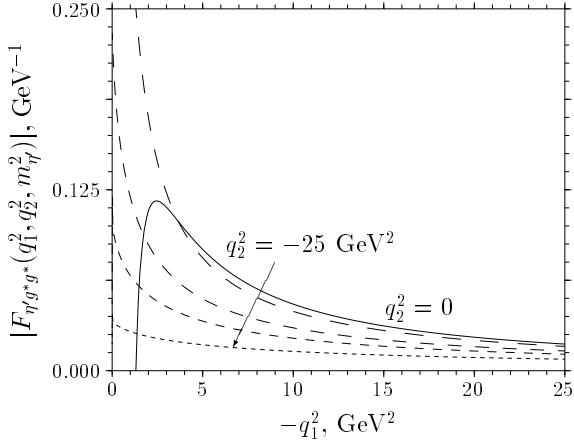,viewport=125 445 460 715,width=.45\textwidth} 
\psfig{file=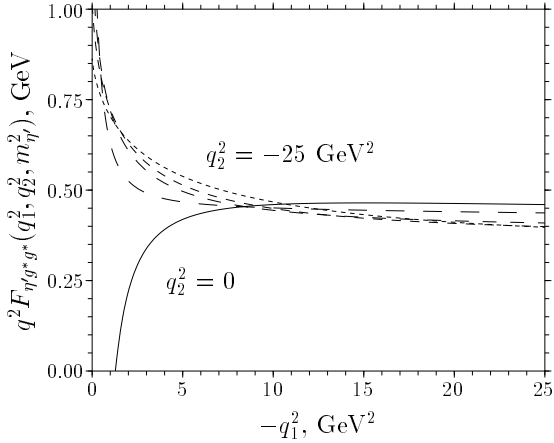,viewport=125 445 460 715,width=.45\textwidth} 
\caption{\label{fig:ff-bl-mp}%
         The $\egg$ vertex functions for the space-like gluon
         virtualities in the forms: $|F_{\eta^\prime g^* g^*}
         (q_1^2, q_2^2, m_{\eta^\prime}^2)|$ (upper figure) 
         and $q^2 F_{\eta^\prime g^* g^*}
         (q_1^2, q_2^2, m_{\eta^\prime}^2)$ (lower figure) 
         at $B^{(q)}_2 = 0.1$ and $B^{(g)}_2 = 3.0$,  
         calculated in the Brodsky-Lepage approach. The legends 
         are as follows: $q_2^2=0$ (solid curve), $q_2^2=-1$ GeV$^2$ 
         (long-dashed curve), $q_2^2=-5$ GeV$^2$ (medium-dashed curve), 
         $q_2^2=-10$ GeV$^2$ (short-dashed curve), $q_2^2=-25$ GeV$^2$ 
         (dotted curve).}
\end{figure}
%
%
%
%
\begin{figure}[tb]
%
\psfig{file=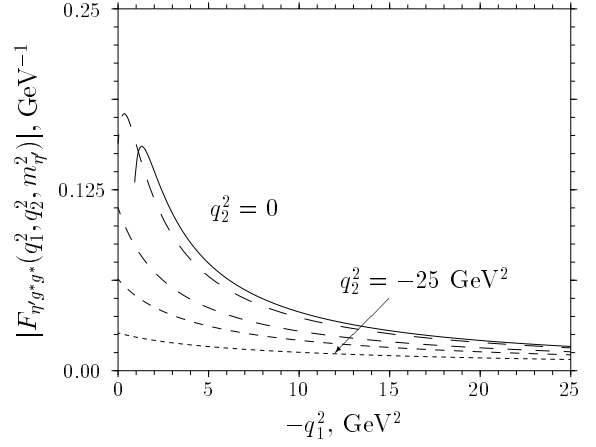,viewport=125 445 460 715,width=.45\textwidth} 
\psfig{file=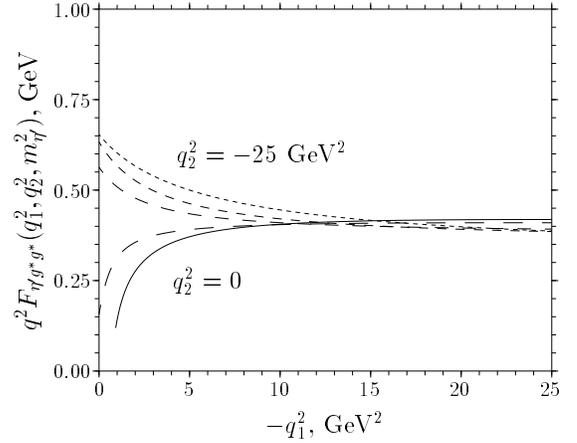,viewport=125 445 460 715,width=.45\textwidth} 
\caption{\label{fig:ff-jk-mp}%
         The $\egg$ vertex functions for the space-like gluon
         virtualities in the forms $|F_{\eta^\prime g^* g^*}
         (q_1^2, q_2^2, m_{\eta^\prime}^2)|$ (upper figure) and
         $q^2 F_{\eta^\prime g^* g^*} (q_1^2, q_2^2, m_{\eta^\prime}^2)$ 
         (lower figure) with $B^{(q)}_2 = 0.1$ and $B^{(g)}_2 = 3.0$,  
         calculated in the mHSA formalism. Legends are the same as in
         Fig.~\ref{fig:ff-bl-mp}.}
\end{figure}
%
%

We shall now present the effect of the transverse   
momenta on the vertex functions.
To that end, the $\egg$ vertex functions 
$|F_{\eta^\prime g^* g^*} (q_1^2, q_2^2, m_{\eta'}^2)|$ and 
$q^2 \,  F_{\eta^\prime g^* g^*} (q_1^2, q_2^2, m_{\eta'}^2)$ 
are presented in the Brodsky-Lepage approach  in Fig.~\ref{fig:ff-bl-mp},
for the space-like gluon virtualities with $q^2 = q_1^2 + q_2^2$ and no
transverse momentum effects taken into account. The corresponding vertex
functions in the mHSA formalism are shown in Fig.~\ref{fig:ff-jk-mp}, in
which transverse momentum
effects are included as discussed earlier. The various curves shown  
correspond to the following values of the second gluon virtuality: 
$q_2^2 = 0$, $-1$, $-5$, $-10$, and $-25$~GeV$^2$. We have fixed 
$B^{(q)}_2 = 0.1$ and $B^{(g)}_2 = 3.0$ from the analysis of
the $F_{\eta^\prime \gamma^* \gamma}$ form factor.
Both approaches give the same asymptotic behavior at large values of
the mass scale~$Q^2$ of the form: $|F_{\egg}| \simeq 0.4~{\rm GeV}/Q^2$.
In the region of small ~$Q^2$, the mHSA-prescription modifies the
behavior of the vertex functions as it decreases the gluonic contribution. 

%
%
\begin{figure}[tb]
%
\psfig{file=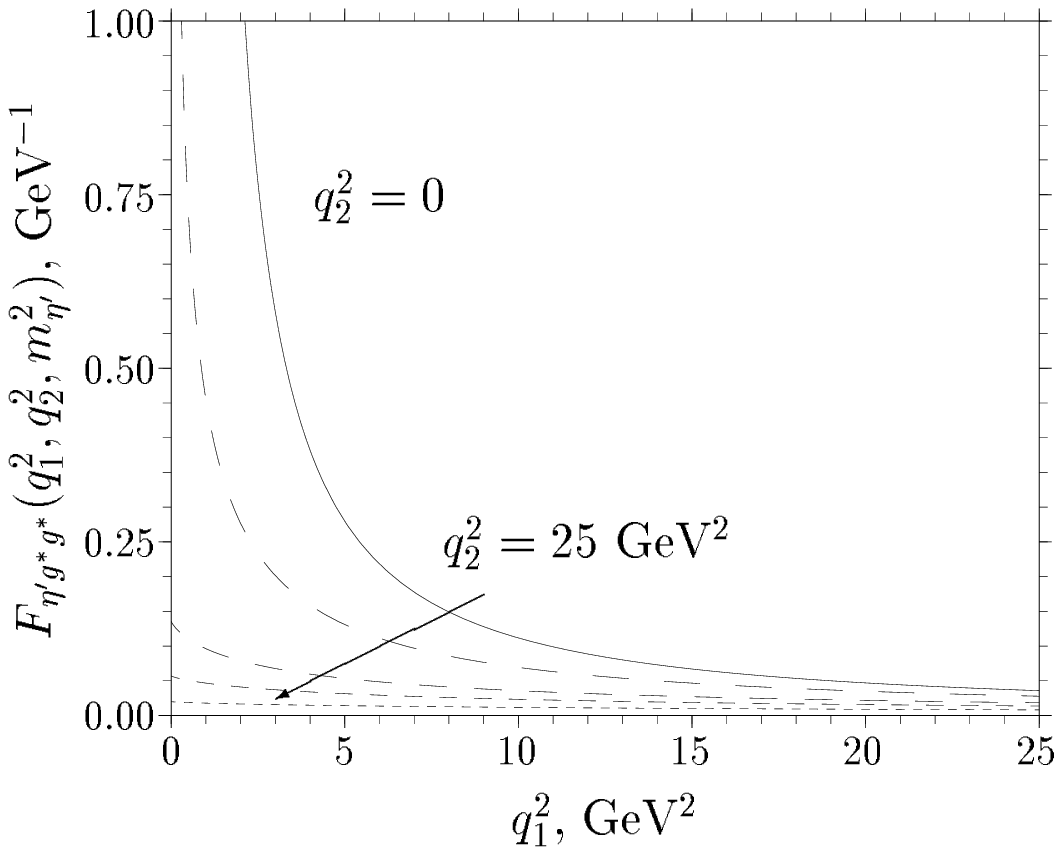,viewport=130 450 460 710,width=.45\textwidth} 
\psfig{file=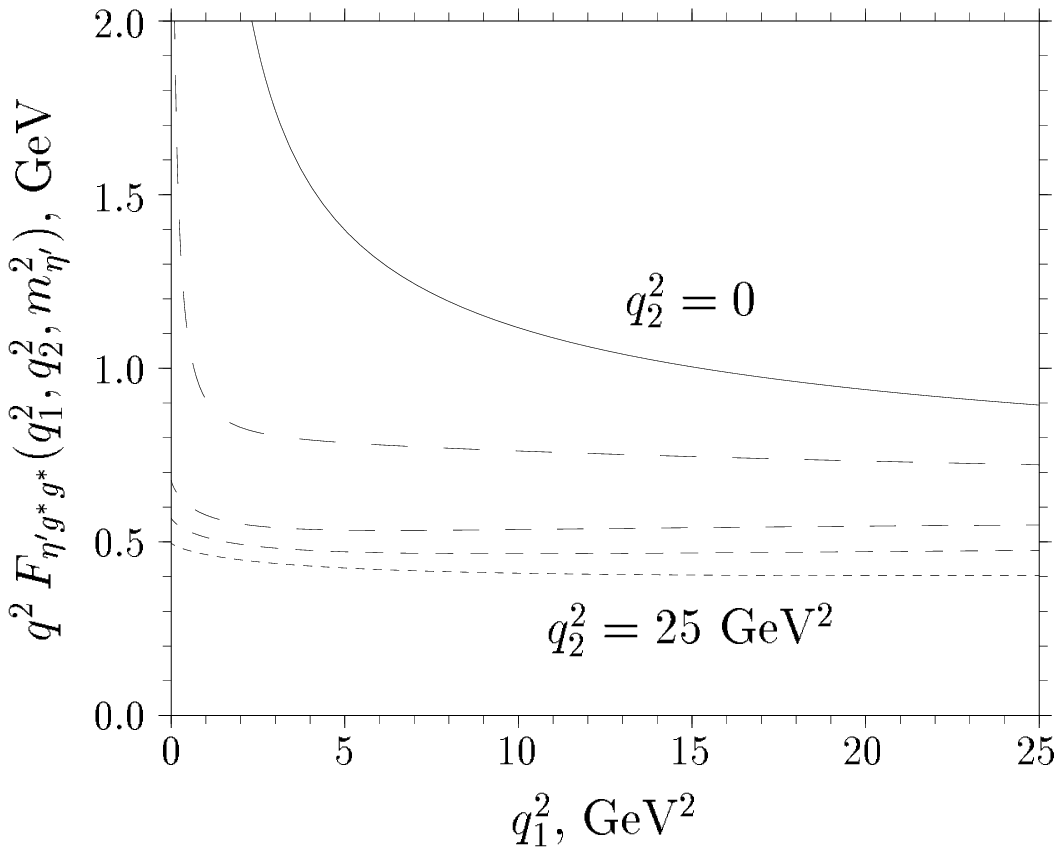,viewport=130 450 460 710,width=.45\textwidth}
\caption{\label{fig:ff-bl-p}%
         The $\eta^\prime g^* g^*$ vertex functions 
         $F_{\eta^\prime g^* g^*} (q_1^2, q_2^2, m_{\eta^\prime}^2)$
         and $q^2 F_{\eta^\prime g^* g^*} 
         (q_1^2, q_2^2, m_{\eta^\prime}^2)$ for 
         time-like gluon virtualities with $B^{(q)}_2 = 0.1$ 
         and $B^{(g)}_2 = 3.0$ in the Brodsky-Lepage approach.
         The legends are as
         follows: $q_2^2=0$ (solid curve), $q_2^2=1$ GeV$^2$ (long-dashed
         curve), $q_2^2=5$ GeV$^2$ (medium-dashed curve), $q_2^2=10$
         GeV$^2$ (short-dashed curve), $q_2^2=25$ GeV$^2$ (dotted curve).}
\end{figure}
%
%

\begin{figure*}[tb]  
%
\centerline{
\psfig{file=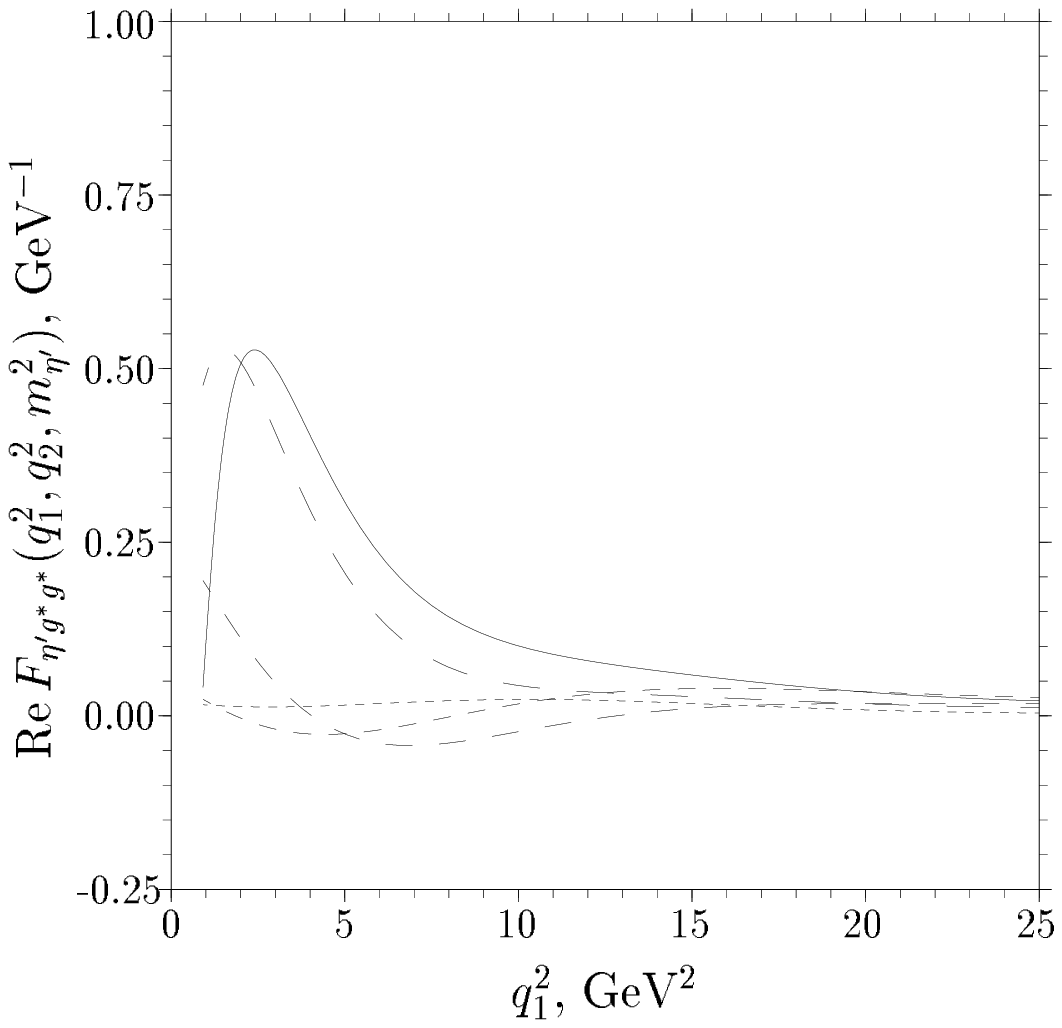,viewport=130 360 460 665,width=.45\textwidth}
\psfig{file=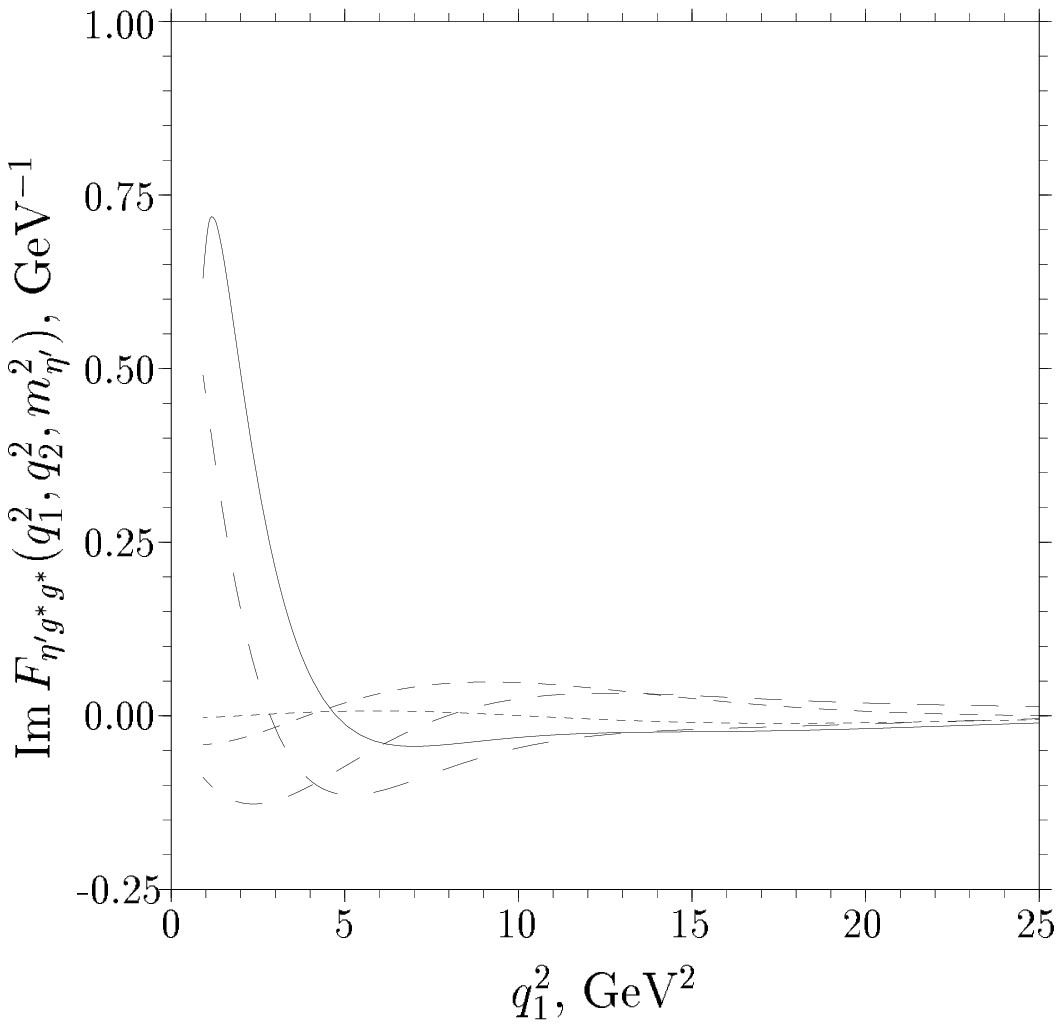,viewport=130 360 460 665,width=.45\textwidth}}
\centerline{ 
\psfig{file=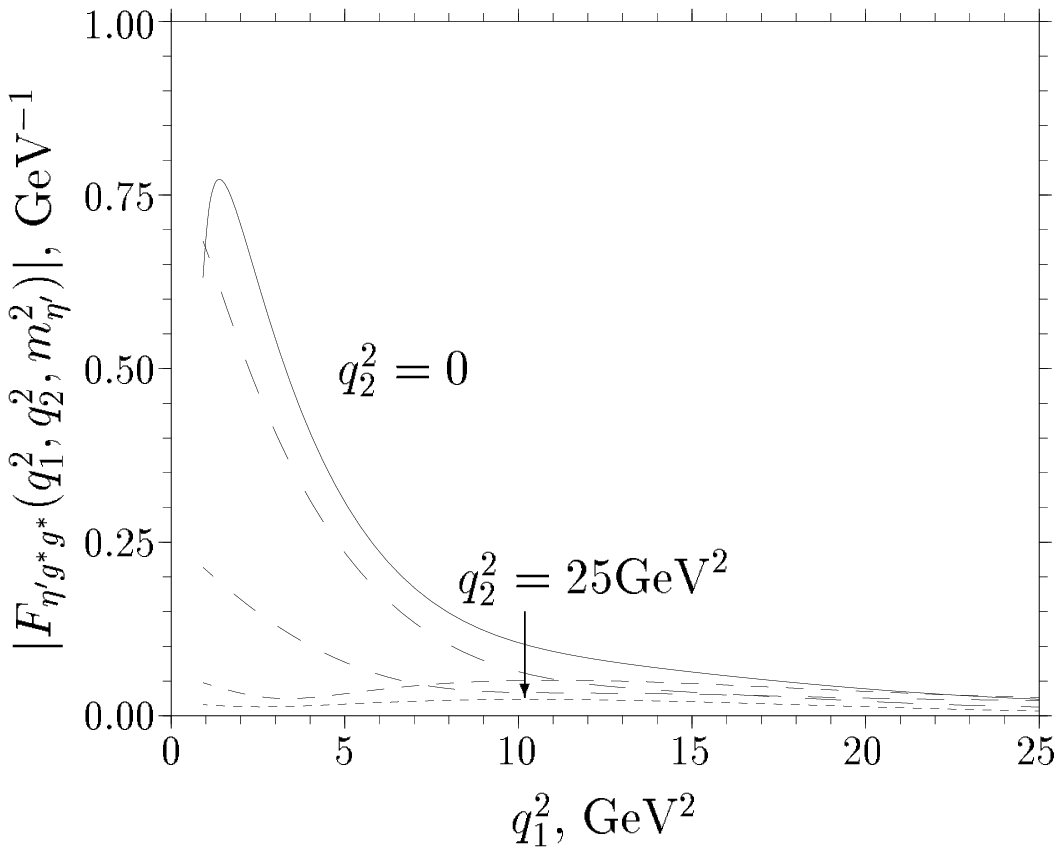,viewport=130 360 460 665,width=.45\textwidth}
\psfig{file=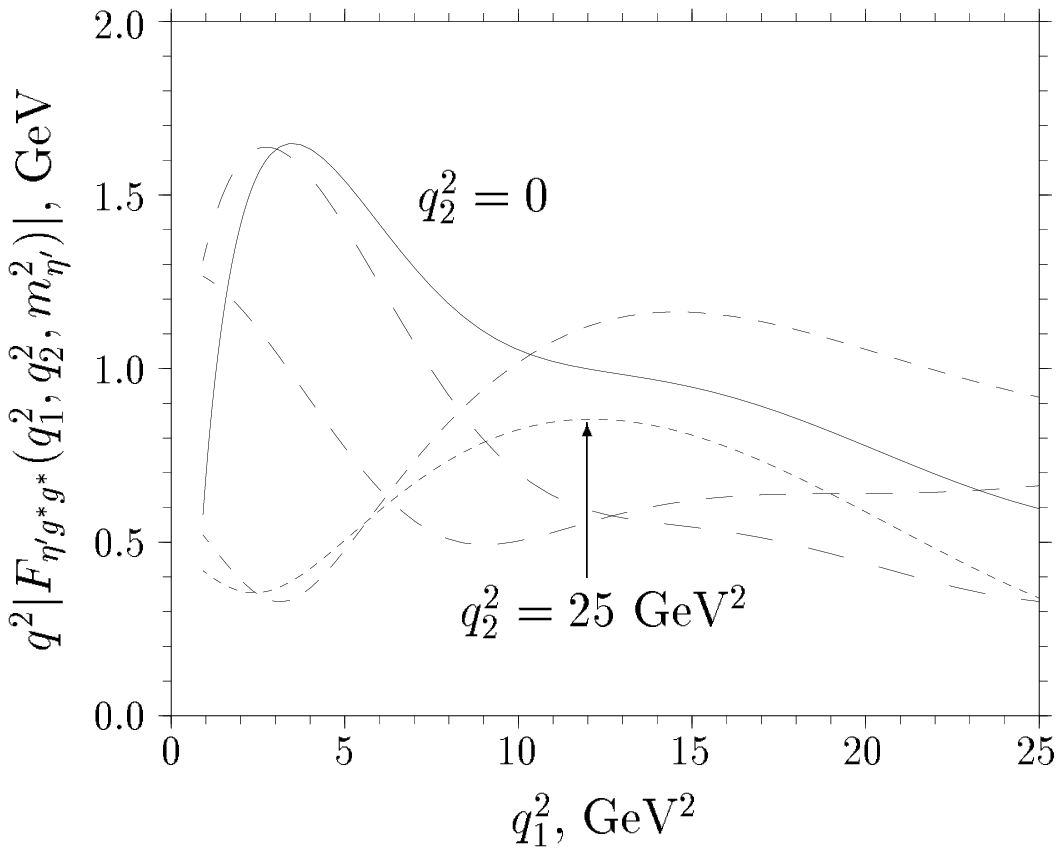,viewport=130 360 460 665,width=.45\textwidth}}
\caption{\label{fig:ff-jk-p}%
         The $\egg$ vertex
         $F_{\eta^\prime g^* g^*} (q_1^2, q_2^2, m_{\eta^\prime}^2)$
         in the mHSA formalism, 
         with the time-like gluon virtualities, $B^{(q)}_2 = 0.1$
         and $B^{(g)}_2 = 3.0$ where $q^2 = q_1^2 + q_2^2$. 
         Legends are the same as in Fig.~\ref{fig:ff-bl-p}.}
\end{figure*}

The results for the $\egg$ vertex with the 
time-like gluon virtualities ($q_1^2 > 0$, $q_2^2 >0$) in the forms     
$F_{\eta^\prime g^* g^*} (q_1^2, q_2^2, m_{\eta^\prime}^2)$ and 
$q^2 \, F_{\eta^\prime g^* g^*} (q_1^2, q_2^2, m_{\eta^\prime}^2)$  
with $B^{(q)}_2 = 0.1$ and $B^{(g)}_2 = 3.0$ in the Brodsky-Lepage 
approach and mHSA are presented in Figs.~\ref{fig:ff-bl-p}
and \ref{fig:ff-jk-p}, 
respectively. We recall that in the considered region of the virtualities,  
an imaginary part appears in the $\egg$ vertex in the mHSA formalism 
in contrast to the Brodsky-Lepage scheme in which the $\egg$ vertex is
defined as real by prescription (see Sec.~\ref{sec:Sudakov}). In both
approaches, 
the absolute values of the function describing the $\egg$ vertex decrease
with increase of the  
second gluon virtuality,~$q_2^2$. When the virtualities of both 
the gluons are  close to the $\eta^\prime$-meson mass squared an 
enhancement of the vertex function is observed in both approaches. 
Taking into account the transverse momentum distribution of partons
inside the $\eta^\prime$-meson in the mHSA, the singularity in the form
factor obtained in the Brodsky-Lepage approach is replaced by the 
resonance-like behavior. 
Nevertheless, qualitatively the behavior of the vertex function in 
the Brodsky-Lepage approach (upper plot in Fig.~\ref{fig:ff-bl-p}) 
and in the mHSA (left-down plot in Fig.~\ref{fig:ff-jk-p}) is similar.   
%
%
\hspace{90mm}
%
%

\begin{figure}[tb]  
%
\psfig{file=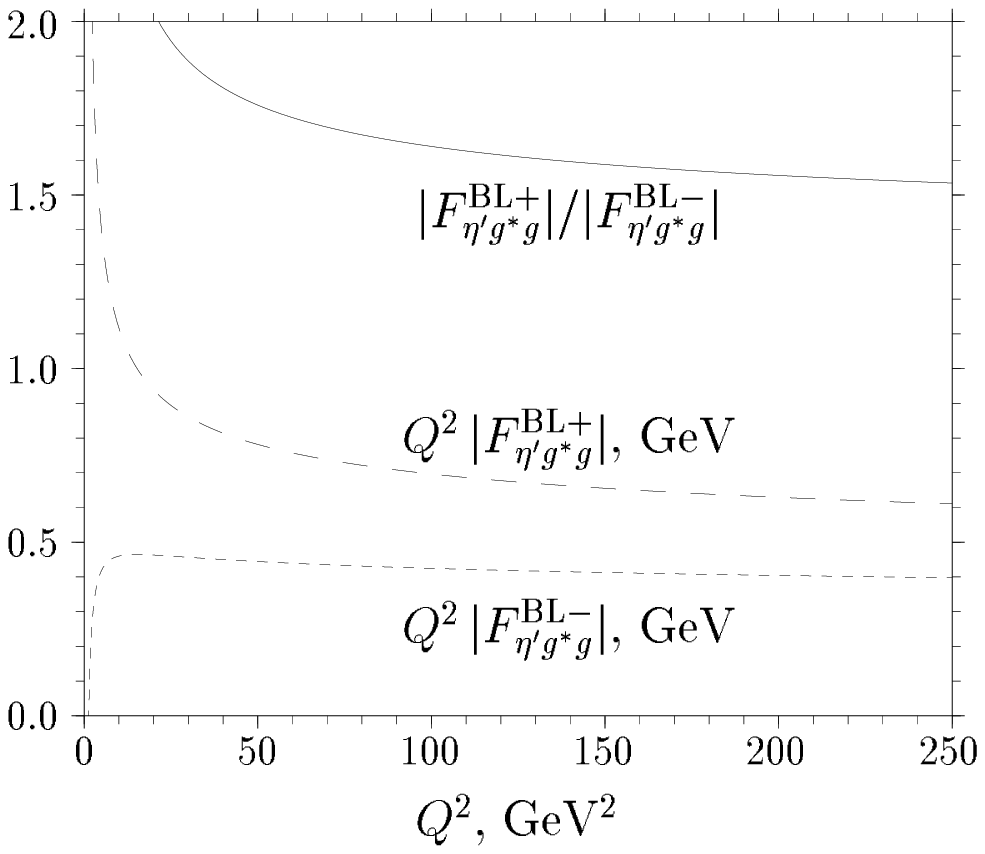,viewport=135 400 460 665,width=.45\textwidth} 
\psfig{file=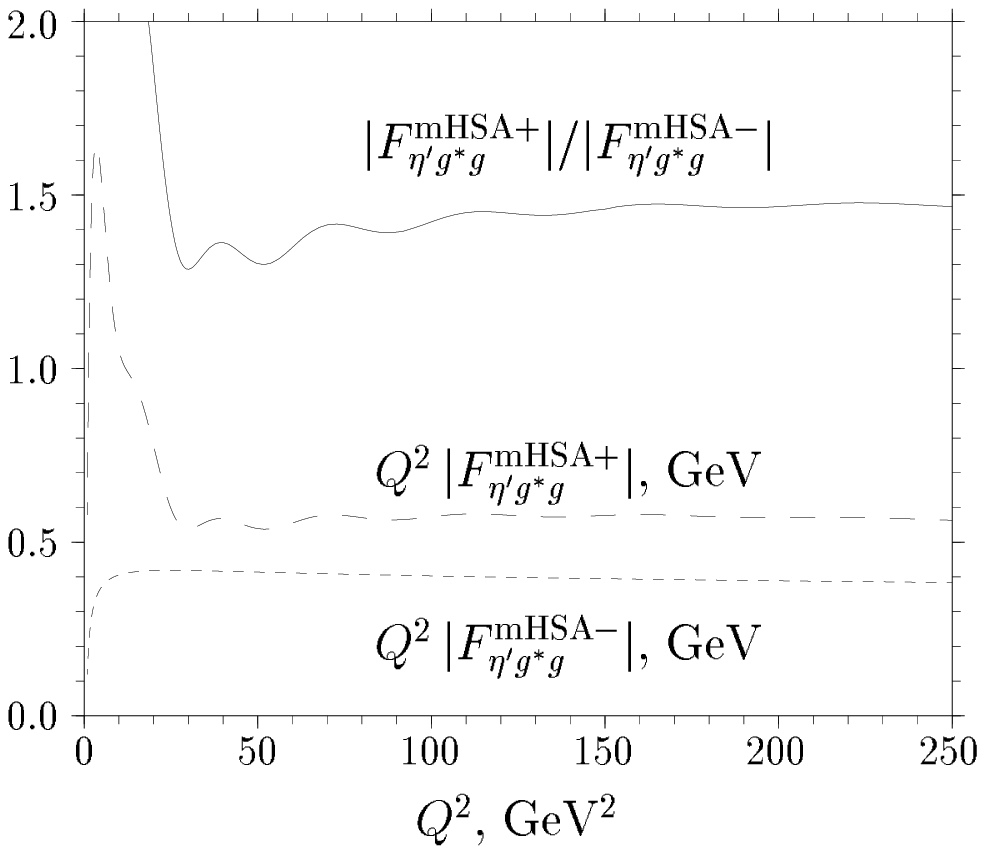,viewport=135 400 460 665,width=.45\textwidth}
\caption{\label{fig:ff-asympt}%
         The large~$Q^2$ asymptotics of the vertex functions 
         $F_{\eta' g^* g} (q_1^2, 0, m_{\eta^\prime}^2)$
         with the time-like ($+$) and space-like ($-$) gluon virtualities 
         with $B^{(q)}_2 = 0.1$ and $B^{(g)}_2 = 3.0$ in the
         Brodsky-Lepage (BL) and the mHSA approaches.} 
\end{figure}

The typical values of the gluon virtualities at which the vertex function 
is well described by its asymptotic behavior $F_{\eta^\prime g^* g^*} 
\sim 1 / Q^2$ can be determined from Fig.~\ref{fig:ff-asympt} in both 
the Brodsky-Lepage and mHSA approaches for the case when one of the 
gluons is on the mass shell. It is seen that for large~$Q^2$ the  
vertex functions in both of these approaches are well correlated. 
In the case of the negative gluon virtualities, the function
$F_{\eta^\prime g^* g^*}$ reaches its asymptotic
form at smaller values of $Q^2$ ($Q^2 \sim 5-10$~GeV$^2$) than in the case 
of positive virtualities ($Q^2 \sim 50-100$~GeV$^2$).  
%
%
\hspace{90mm}
%
%

\section{\label{sec:Approximate}%
         Interpolating Formulae for the $\eta^\prime g^* g^*$ Vertex}

For the applications in various decay and production processes it is
useful to find some approximate
formulae for the vertex functions which are simple and can be used over a
large domain of the gluon virtualities. We recall that
Brodsky and Lepage~\cite{BL81} presented an
approximate form for the $\pi-\gamma$ transition form factor
which interpolates between the PCAC value and the QCD prediction
in the large~$Q^2$ region. Subsequently, in Ref.~\cite{FK98-2}, this
form was extended to the case of the $\eta^\prime-\gamma$ transition form
factor. Very much along the same lines, a similar expression can be
written for the $\eta^\prime g^* g^*$ vertex:
\begin{equation}
F^{\rm BL}_{\egg} (q^2, \omega) = 4 \pi \alpha_s (Q^2) \,
\frac{2 \sqrt{3} f_\pi D (q^2, \omega)}
     {\sqrt{3} \, q^2 - 8 \pi^2 f_\pi^2 D (q^2, \omega)} ,
\label{eq:BL-appr}
\end{equation}
where the largest energy scale parameter $Q^2 = q^2$ for the time-like
total gluon virtuality and $Q^2 = - q^2$ for the space-like one.
We introduce also the following function:
\begin{eqnarray}
&& D (q^2, \omega) = f_0 (\omega) + \frac{q^2}{Q^2} \, g_2 (\omega) 
\label{eq:D-function} \\
&& \times 
\left [ 16 B^{(q)}_2
\left ( \frac{\alpha_s (Q^2)}{\alpha_s (\mu_0^2)} \right )^{\frac{48}{81}}
\! \! \!
+ 5 B^{(g)}_2 \left ( \frac{\alpha_s (Q^2)}{\alpha_s (\mu_0^2)}
\right )^{\frac{101}{81}}
\right ] .
\nonumber 
\end{eqnarray}
When one of the gluons is on the mass shell ($q_2^2 = 0$) the asymptotic
functions $f_0 (\omega)$ and $g_2 (\omega)$ have the following values:
$f_0 (1) = 1$ and $g_2 (1) = 1/6$.
We do not include the next-to-leading quark contribution which is
strongly suppressed. In the numerical analysis the usual prescription for
the
evaluation of strong coupling should be used: it is estimated using the
value of the largest scale parameter of the problem. For the anomaly,
with real gluons, we use $Q^2 = m_{\eta^\prime}^2$. The above expression
reproduces both the anomaly value and the large~$Q^2$
asymptotics of the vertex function:
\begin{eqnarray}
\left . F_{\egg} (Q^2, \omega) \right |_{Q^2 \to 0} & = &
4 \pi \alpha_s (m_{\eta^\prime}^2) \, \frac{\sqrt 3}{4 \pi^2 f_\pi} ,
\label{eq:asymp-0} \\
\left . F_{\egg} (Q^2, \omega) \right |_{Q^2 \to \infty}
& = & 4 \pi \alpha_s (Q^2) \, \frac{2 f_\pi D (q^2, \omega)}{q^2} .
\qquad 
\label{eq:asymp-inf}
\end{eqnarray}
Presented in the form given in Eq.~(\ref{eq:BL-appr}), the vertex
$F^{\rm BL}_{\egg} (q^2, \omega)$ is a smooth function in
the space-like region of the gluon virtualities but has a pole at
$q^2 = 8 \pi^2 f_\pi^2 D(q^2, \omega) / \sqrt 3$ in the time-like
region. A similar behavior for the $\egg$ vertex in the time-like
region was obtained by Kagan and Petrov~\cite{TFF-1} from
the evaluation of the triangle diagram.

The mHSA approach removes the unphysical singularity from
the vertex in the time-like region of the gluon virtualities
but the vertex gets an imaginary part. In this case the real
part of the approximate formula interpolates the anomaly value and
the large~$Q^2$ asymptotics, while the imaginary part goes to zero
as $Q^2 \to 0$. It is possible to get the approximate formula
$F^{\rm mHSA}_{\egg} (Q^2, \omega)$ without the pole behavior in the
time-like region of the gluon virtualities, if one takes into account
the corrections of order $1 / Q^{5/2}$~(\ref{eq:FFq-mHSA-asymp}) 
to the vertex function. In this case the functional 
form defined in Eq.~(\ref{eq:BL-appr}) is still valid but with the
replacement: 
\begin{eqnarray} 
D (Q^2, \omega) & \to & {\cal D} (Q^2, \omega) = D (Q^2, \omega) + 
\frac{2 \pi i}{\omega} \, y_+ y_- 
\label{eq:Dc-function} \\ 
& \times &
(1 - y_+ y_-) \, 
\left [ H_0^{(2)} (Q_\Lambda y_+) - H_0^{(2)} (Q_\Lambda y_-) \right ] ,  
\nonumber 
\end{eqnarray}
where $y_\pm$ is defined by Eq.~(\ref{eq:FFq-mHSA-asymp}) and 
the Hankel function $H_0^{(2)} (Q_\Lambda y_\pm)$ is assumed 
in the limit of the large asymptotics of it argument.
As the function ${\cal D} (Q^2, \omega)$ is complex, 
the vertex $F^{\rm mHSA}_{\egg} (Q^2, \omega)$ becomes a complex
function as well. In the limit of large $Q^2$, Eq.~(\ref{eq:BL-appr})
with the replacement~(\ref{eq:Dc-function}) reproduces the
asymptotics~(\ref{eq:FFq-mHSA-asymp}) of the vertex function. The real
part of Eq.~(\ref{eq:BL-appr}) gives the value determined by the
anomaly~(\ref{eq:asymp-0})
in the limit $Q^2 \to 0$. As for the imaginary part of the vertex
function,
it goes to zero as ${\rm Im} F^{\rm mHSA}_{\egg} (Q^2, \omega) \sim Q^4$ 
in the limit of small virtualities according to the approximate
formula~(\ref{eq:BL-appr}).

\section{\label{sec:concl}Summary}

In the present paper we have analyzed the $\egg$ vertex, 
$F_{\egg} (q_1^2, q_1^2, m_{\eta^\prime}^2)$, for 
off-mass-shell gluons in the hard scattering approach. The
$\eta^\prime$-meson wave-function is evolved using the
evolution equations for the quark and gluonic components. 
In difference to Ref.~\cite{Muta} it is shown that, within the possible 
variation of the parameters $B^{(q)}_2$ and $B^{(g)}_2$
of the $\eta^\prime$-meson wave-function, the gluonic contribution
can not be ignored. To further quantify this, one needs to know the
non-perturbative parameters entering in the evolution of the
$\eta^\prime$-meson wave-function. Using the maximum values of these 
parameters $|B^{(q)}_2| = 0.1$ and $|B^{(g)}_2| = 9.0$, we find that the 
gluonic contribution is comparable to the quark contribution, leading to 
a sizable (almost a factor 2) enhancement of the $\eta^\prime g^* g^*$
vertex. The magnitude of the vertex is not very sensitive to the 
value of the parameter~$B^{(q)}_2$, and even a factor three to five
reduction from its default value used here does not change the vertex
function appreciably. However, the vertex function is rather sensitive to
the input value of the parameter~$B^{(g)}_2$. We have explicitly shown
this dependence in our work. Hence, it is important to constrain or
measure this quantity for more definitive predictions. However, we find  
that even with a significantly reduced value of~$B^{(g)}_2$, say for
$|B^{(g)}_2| = 3.0$, the gluonic correction is of the order of a few tens
percent, and hence it must be included in all perturbative treatments of
the $\eta^\prime g^* g^*$ vertex function.

We have obtained analytic expressions for the vertex describing the
$\eta^\prime$-meson transition into two gluons with arbitrary
virtualities in the Brodsky-Lepage approach in which the transverse
momentum dependence of the partons inside the $\eta^\prime$-meson is
ignored. The appearance of a singularity in the
region of the $\eta^\prime$-meson mass indicates that this
approach is valid only in the asymptotic region for the total gluon
virtuality, $q^2 = q_1^2 + q_2^2$, i.e., for $|q^2| \gg
m_{\eta^\prime}^2$.  In this region the $\eta^\prime g^* g^*$ vertex has
the usual behavior as the pseudoscalar meson form factors: $F_{\egg} \sim
1 / q^2$. We have compared our results with some of the existing 
parametrizations of the $\egg$ vertex,
$F_{\egg} (q_1^2,0,m_{\eta^\prime}^2)$, in the time-like region. 
Corresponding results for the vertex $F_{\egg}(q_1^2,0,m_{\eta^\prime}^2)$
in the space-like region of $q_1^2$ are also presented, and the behavior
of the $\egg$ vertex is found to be close to that of the electromagnetic
transition form factor of the $\eta^\prime$-meson~-- an information which 
has been used to fix the parameters $B^{(q)}_2$ and $B^{(g)}_2$ in the 
evolution of the $\eta^\prime$-meson wave-function.

 The mentioned singularity in the Brodsky-Lepage approach can be
circumvented in the modified hard scattering approach (mHSA), in which 
the transverse momentum dependence of the hard scattering amplitude as 
well as the transverse momentum distribution
and soft-gluon emission (the Sudakov effect) in the
$\eta^\prime$-meson wave-function are taken into account. 
Analytic properties of the $\egg$ vertex are reviewed  to obtain the
correct expression in the time-like 
region of the gluon virtualities. It is shown that due to the 
$i \epsilon$ prescription of the propagators, the $\egg$ vertex 
acquires an additional phase factor in comparison with the 
space-like expression. We present the resulting $\egg$ vertex in the mHSA
formalism for the space-like and time-like virtualities and work out
the large $Q^2$ asymptotics.
 Numerical analysis shows that in the modified hard
scattering approach  the $\egg$ vertex in the time-like region of the
gluon virtualities reaches the asymptotic form at rather large values,
$Q \gtrsim 10$~GeV, while in the space-like
region, the corresponding vertex  can be used in its asymptotic form
already at $Q \sim 3-5$~GeV. Finally, we have provided simple
interpolating formulae for the vertex $F_{\egg}(q^2,\omega)$ for
the time-like and space-like gluon virtualities, which reproduce the
anomaly (for on-shell gluons) and the asymptotic form, determined in the
hard scattering approach. The computed vertex $F_{\egg}(Q^2,\omega)$ has
ready applications in a large number of decays and production processes
involving an $\eta^\prime$-meson.

\begin{acknowledgments}

We would like to thank Andrei Belitsky, Vladimir Braun, John Collins,
Markus Diehl, Eduard Kuraev, and Lev Lipatov for helpful discussions.
We thank Taizo Muta for correspondence on Ref.~\cite{Muta}, and Rainer
Jakob, Peter Kroll, Hsiang-Nan Li, and George Sterman for clarifying the
literature  on the Sudakov form factor of the pion. We thank Dmitrii
Ozerov for helpful discussions on numerical computations. 
A.P. would like to thank the DESY theory group for its hospitality in
Hamburg where the major part of this work was done. The work of A.P. is
partially supported by the Russian Foundation for Basic Research under
Grant No. 98-02-16694, and in part by the German Academic Exchange
Service DAAD.  

\end{acknowledgments}

\appendix 

\section{\label{app:evol-eqns}%
         Solutions of Evolution Equations}

The solutions for the quark and gluonic wave-functions of
the color- and SU(3)$_F$ flavor-singlet pseudoscalar meson  
were obtained in 
Ref.~\cite{Ohrndorf:1981uz,Shifman:1981dk,Baier:1981pm}. 
Their general forms are:
\begin{widetext}
\begin{eqnarray}
\hspace*{-20mm} &&
\phi^{(q)}(x,Q) = 6 x \bar x
\left \{ 1 + \!\!\!
\sum_{{\rm even} \, n \ge 2} \left [ B^{(q)}_n
\left ( \frac{\alpha_s (\mu_0^2)}{\alpha_s (Q^2)} \right )^{\gamma_+^n}
\!\!\! + \rho^{(g)}_n B^{(g)}_n
\left ( \frac{\alpha_s (\mu_0^2)}{\alpha_s (Q^2)} \right )^{\gamma_-^n}
\right ] C^{3/2}_n (x - \bar x)
\right \} , 
\label{eq:qef-gen} \\
\hspace*{-20mm} &&
\phi^{(g)}(x,Q) = x \bar x 
\sum_{{\rm even} \, n \ge 2}
\left [ \rho^{(q)}_n B^{(q)}_n
\left ( \frac{\alpha_s (\mu_0^2)}{\alpha_s (Q^2)} \right )^{\gamma_+^n}
\!\!\! + B^{(g)}_n
\left ( \frac{\alpha_s (\mu_0^2)}{\alpha_s (Q^2)} \right )^{\gamma_-^n}
\right ] C^{5/2}_{n - 1} (x - \bar x) .
\label{eq:gef-gen}
\end{eqnarray}
\end{widetext}
Here, $Q^2$ is the scale of the hard process, $\mu_0 \simeq 0.5$~GeV 
is the typical hadronic energy scale below which no perturbative
evolution takes place, $x$ and $\bar x = 1 - x$ 
are the momentum fractions of partons inside the pseudoscalar meson,
$C^\nu_n (x)$ are the Gegenbauer polynomials of the order~$n$ with the
index~$\nu$~\cite{GR}. When the index is semi-integer, $\nu = m + 1/2$, 
the Gegenbauer polynomials can be defined by the following relation:
\begin{equation}
C^{m + 1/2}_{n - m} (x) = \frac{2^m m!}{2^n n! (2 m)!}
\left ( \frac{d}{d x} \right )^{m + n} (x^2 - 1)^n .
\label{eq:GP-def}
\end{equation}
In our analysis we are interested in the polynomials of the first
three orders:
\begin{eqnarray}
C^{m + 1/2}_0 (x) & = & 1 ,
\nonumber \\
C^{m + 1/2}_1 (x) & = & (2 m + 1) x ,
\label{eq:GP-first} \\
C^{m + 1/2}_2 (x) & = & \frac{1}{2} (2 m + 1)
\left [ (2 m + 3) x^2 - 1 \right ] .
\nonumber
\end{eqnarray}
Hence, $C^{3/2}_2 (x) = \frac{3}{2} (5 x^2 - 1)$ and 
$C^{5/2}_1 (x) = 5 x$. The differential equation:
\begin{equation}
\frac{d}{dx} C^\nu_n (x) = 2 \nu  \, C^{\nu + 1}_{n - 1} (x) ,  
\label{eq:Diff-eq}
\end{equation}
allows to connect the coefficients of the highest powers of these 
polynomials as ${}^{3/2}_{\;\; n} a_n = (3/n) \, {}^{5/2}_{n-1}
a_{n-1}$, 
where the following representation for the polynomials is assumed:
\begin{equation} 
C^\nu_n (x) = {}^\nu_n a_n \, x^n + {}^\nu_n a_{n-1} \, x^{n-1} + 
\ldots + {}^\nu_n a_0 .
\label{eq:GP-representation}
\end{equation}
Eqs.~(\ref{eq:qef-gen}) and~(\ref{eq:gef-gen}) contain the set
of parameters called~$\gamma_{\pm}^n$, defined as:
\begin{equation}
\gamma_{\pm}^n =  \frac{1}{2} \,
\left [ \gamma_{QQ}^n + \gamma_{GG}^n
\pm \sqrt{(\gamma_{QQ}^n - \gamma_{GG}^n)^2 +
          4 \gamma_{QG}^n \gamma_{GQ}^n} \,
\right ],
\label{eq:AD-rotate}
\end{equation}
where the anomalous dimensions~$\gamma_{ij}^n$ 
are~\cite{Shifman:1981dk,Baier:1981pm,Belitsky}:
\begin{eqnarray}
\gamma_{QQ}^n & = & \frac{C_F}{\beta_0}
\left [ 3 + \frac{2}{(n + 1)(n + 2)} -
        4 \sum_{j = 1}^{n + 1} \frac{1}{j} \right ] ,
\nonumber \\
\gamma_{GQ}^n & = & \frac{C_F}{\beta_0} \, 
\frac{n (n + 3)}{3 (n + 1) (n + 2)} ,
\label{AD-matrix} \\
\gamma_{QG}^n & = & \frac{n_f}{\beta_0} \,
\frac{12}{(n + 1)(n + 2)} ,
\nonumber \\
\gamma_{GG}^n & = & \frac{N_c}{\beta_0}
\left [ \frac{8}{(n + 1)(n + 2)} -
        4 \sum_{j = 1}^{n + 1} \frac{1}{j} \right ] + 1 . 
\nonumber 
\end{eqnarray}
Here, $N_c$ is the number of colors, $C_F = (N_c^2 - 1)/ (2 N_c)$ is the
eigenvalue of the Casimir operator in the fundamental representation
of the SU($N_c$) group, $n_f$ is the number of active quarks,
$\beta_0 = 11 - 2 n_f/3$ is the one-loop $\beta$-function coefficient, 
and $n \ge 1$. Note that in Ref.~\cite{Ohrndorf:1981uz} the
factors~$3/n$ and~$n/3$ were not included in the nondiagonal anomalous
dimensions~$\gamma_{QG}^n$ and~$\gamma_{GQ}^n$ presented above, as well
as a factor~2 was missed in~$\gamma_{QG}^n$.
 The detailed discussion of these anomalous dimensions in 
the one- and two-loop approximation can be found in Ref.~\cite{Belitsky}.
These $\gamma$'s allow to define the $\rho$~parameters entering in
Eqs.~(\ref{eq:qef-gen}) and~(\ref{eq:gef-gen}): 
\begin{equation}
\rho^{(q)}_n = 6 \, \frac{\gamma_+^n - \gamma_{QQ}^n}{\gamma_{GQ}^n} ,
\qquad
\rho^{(g)}_n = \frac{1}{6} \,
\frac{\gamma_{GQ}^n}{\gamma_-^n - \gamma_{QQ}^n} .
\label{eq:rho's}
\end{equation}
The numerical values of the $\gamma$'s and $\rho$'s needed in the
numerical analysis are given below for QCD with $n_f=3$:
\begin{eqnarray}
\begin{array}{ll}
\gamma_{QQ}^2 = - \frac{50}{81} , \qquad &
\gamma_{GQ}^2 = \frac{10}{243} , \\[1mm]
\gamma_{GG}^2 = - \frac{11}{9} , \qquad &
\gamma_{QG}^2 = \frac{1}{3} , \\[1mm]
\gamma_+^2 \simeq - \frac{48}{81} , \qquad &
\gamma_-^2 \simeq - \frac{101}{81} , \\[1mm]
\rho^{(q)}_2 \simeq \frac{16}{5} , \qquad &
\rho^{(g)}_2 \simeq - \frac{1}{90} .
\end{array}
\label{eq:num-values}
\end{eqnarray}
The parameters $B^{(q)}_n$ and $B^{(g)}_n$ are not determined
by perturbative QCD and are treated as free parameters, to be
fixed by data, for example, from the $\eta^\prime - \gamma$
transition.

\section{\label{app:J-func}%
         The Function ${\rm J} (\omega, \eta)$}

The integral ${\rm J}(\omega,\eta)$ used in the main text is defined as
follows: 
\begin{equation}
{\rm J} (\omega, \eta) \equiv \int\limits_0^1 
\frac{dx}{1 + \omega (x - \bar x) - 2 x \bar x \eta + i \epsilon} . 
\label{eq:J-func}
\end{equation}
It is easy to see that this function is symmetric on its first argument: 
${\rm J} (- \omega, \eta) = {\rm J} (\omega, \eta)$. 

The result of the integration depends on the correlation between~$\omega$ 
and~$\eta$ and is the following:  
\begin{widetext}
\begin{equation}
{\rm J} (\omega, \eta) = 
\left \{ 
\begin{array}{ll} 
\frac{1}{\sqrt{1 - \omega^2 - (1 - \eta)^2}} 
\left [ 
\arctan \frac{\eta + \omega}{\sqrt{1 - \omega^2 - (1 - \eta)^2}} + 
\arctan \frac{\eta - \omega}{\sqrt{1 - \omega^2 - (1 - \eta)^2}}  
\right ], & 
\omega^2 + (1 - \eta)^2 \le 1 ; \\ 
\frac{1}{2 \sqrt{(1 - \eta)^2 - 1 + \omega^2}} \left [ 
\ln \left | 
\frac{1 - \eta + \sqrt{(1 - \eta)^2 - 1 + \omega^2}} 
     {1 - \eta - \sqrt{(1 - \eta)^2 - 1 + \omega^2}} 
\right | - i \pi \, \Theta(\eta) \right ], & 
\omega^2 + (1 - \eta)^2 > 1,   
\end{array}
\right. 
\label{eq:J-func-res}
\end{equation}
where $\Theta (\eta)$ is the unit step function. 
In the limit $\omega \to \pm 1$ this function has a logarithmic  
divergence:  
\begin{equation}
{\rm J} (1 - 2 \varepsilon, \eta) = \frac{1}{2 (1 - \eta)} 
\left [ 1 + \frac{2 \varepsilon}{(1 - \eta)^2} \right ]
\ln \frac{(1 - \eta)^2}{\varepsilon} + 
\frac{\varepsilon}{2 (1 - \eta)^3} 
- \frac{i \pi \, \Theta(\eta)}{2 (1 - \eta)} \, 
\left [ 1 + \frac{2 \varepsilon}{(1 - \eta)^2} \right ] 
+ {\cal O} (\varepsilon^2) . 
\label{eq:J-func-w1}
\end{equation}
Another asymptotics of the function ${\rm J} (\omega, \eta)$
is also used:  
\begin{equation}
{\rm J} \left ( \frac{1}{1 - 2 \varepsilon},
                \frac{\eta}{1 - 2 \varepsilon} \right )
= \frac{1}{2 (1 - \eta)}
\left [ 1 - \frac{2 \varepsilon (1 + \eta^3)}{(1 - \eta)^3} \right ]
\ln \frac{(1 - \eta)^2}{\varepsilon} - 
\frac{\varepsilon (1 - 2 \eta)}{(1 - \eta)^2} 
- \frac{i \pi \, \Theta(\eta)}{2 (1 - \eta)} 
\left [ 1 - \frac{2 \varepsilon (1 + \eta)}{(1 - \eta)^2} \right ]
+ {\cal O} (\varepsilon^2) .
\label{eq:J-func-1w1} 
\end{equation}
At small values of the second argument $\eta \to 0$ this function has 
the following asymptotic expansion: 
\begin{eqnarray}
{\rm J} (\omega, \eta) & \simeq & 
- \frac{\eta}{\omega^2} 
\left [ 
1 + \frac{3 \eta}{2 \omega^2} + 
\frac{\eta^2 (15 - 4 \omega^2)}{6 \omega^4} + 
\frac{5 \eta^3 (21 - 11 \omega^2)}{24 \omega^6}  
\right ] 
\label{eq:J-func-eta0} \\
& + & \frac{1}{2 \omega} 
\left [ \ln \left | \frac{1 + \omega}{1 - \omega} \right | 
- i \pi \Theta (\eta) \right ] 
\left [ 1 + \frac{\eta}{\omega^2} + 
\frac{\eta^2 (3 - \omega^2)}{2 \omega^4} + 
\frac{\eta^3 (5 - 3 \omega^2)}{2 \omega^6} +  
\frac{\eta^4 (35 - 30 \omega^2 + 3 \omega^4)}{8 \omega^8}  
\right ]   
\nonumber 
\end{eqnarray}
\end{widetext}

\end{document}